# The electromagnetic field outside the steadily rotating relativistic uniform system

**Sergey G. Fedosin**

PO box 614088, Sviazeva str. 22-79, Perm, Perm Krai, Russia

E-mail: sergey.fedosin@gmail.com

**Abstract:** Using the method of retarded potentials approximate formulas are obtained that describe the electromagnetic field outside the relativistic uniform system in the form of a charged sphere rotating at a constant speed. For the near, middle and far zones the corresponding expressions are found for the scalar and vector potentials, as well as for the electric and magnetic fields. Then these expressions are assessed for correspondence to the Laplace equations for potentials and fields. One of the purposes is to test the truth of the assumption that the scalar potential and the electric field depend neither on the value of the angular velocity of rotation of the sphere nor on the direction to the point where the field is measured. However, calculations show that potentials and fields increase as the observation point gets closer to the sphere's equator and to the sphere's surface, compared with the case for a stationary sphere. In this case, additions are proportional to the square of the angular velocity of rotation, the square of the sphere's radius and inversely proportional to the square of the speed of light. The largest found relative increase in potentials and fields could reach the value of 4% for the rapidly rotating neutron star PSR J1614-2230, if the star were charged. For a proton, a similar increase in fields on its surface near the equator reaches 54%.

## 1. Introduction

In article [1] it is emphasized that in most cases calculation of the components of electromagnetic field of rapidly changing currents is extremely difficult. Even in simple configurations of moving charges, non-elementary integrals appear that cannot be expressed in terms of simple functions. The simplest example is a current loop, and already here we have to deal with elliptic integrals. To determine the field components, Maxwell equations for the vector potential were integrated in [1] using the Laplace transformations, and the solution was found in the form of a sum with the help of Legendre polynomials for the charged spherical shell during its rotation in different cases, including change in the charge configuration on the surface and accelerated rotation.

The solution for the rotating uniformly charged sphere's surface can be found in [2], where the magnetic field was expressed as a vector in the spherical reference frame. In [3], the vector potential and magnetic field are calculated for a uniformly charged rotating sphere. A more complicated situation, where the matter inside the sphere or cylinder is a conductor and an additional charge appears during rotation from the centripetal force and inertia of electrons, is considered in [4-5].

In [6], rotating cylindrical charge distribution was studied and a solution was obtained for the magnetic and electric fields around the rotating sphere. Then, in [7] a general solution was found for symmetric rotating charge distributions.

In contrast to these works, we consider not just uniformly charged matter distributed inside the sphere or in its shell, but a relativistic uniform system. This means that the matter in the sphere's volume is in equilibrium with the gravitational forces, pressure field and acceleration field, and the charged particles can move chaotically and have the same invariant charge density. If such a system of particles rotates at a certain constant angular velocity, this leads to the corresponding vector potential and magnetic field, which do not depend on time. We will calculate all the components of the electromagnetic field outside the system, including the scalar and vector potentials, electric and magnetic fields. Previously, these quantities were found in [8-12] for the case of a uniform system at rest without rotation, in which the vector potentials are equal to zero.

The study of a rotating relativistic uniform system is important in itself and it is of academic interest from the point of view of developing an ideal model corresponding to the relativistic approach. But there are also a number of physical problems, such as calculating the angular momentum, magnetic moment, relativistic energy of rotating objects, where it is necessary to correctly estimate the contributions of various fields associated with these objects.

As a rule, in articles describing a steadily rotating spherical shell, it is assumed that the electric field outside the sphere does not depend on the angular velocity of rotation. In contrast to this, in [13] it is indicated that there is such a dependence both for the electric and magnetic fields. In [14] this question was considered again and an error in calculations was found in [13], associated with the replacement of the partial time derivative with the total derivative.

To check the assumption about the possible dependence of the fields on the angular velocity of rotation and to estimate the contribution from the particles' motion inside the system, the accuracy of our calculations will be increased up to the terms containing the square and even the third power of the speed of light in the denominator. The method of retarded potentials used for calculations provides the result based on first principles, which reduces possible inaccuracies that appear under additional assumptions.

## 2. Statement of the problem

The standard equations for the electric field strength $\mathbf{E}$, magnetic field induction $\mathbf{B}$ and electromagnetic field potentials in the framework of the special theory of relativity have the following form:

$$\nabla \cdot \mathbf{E} = \frac{\gamma \rho_{0q}}{\varepsilon_0}, \qquad \nabla \times \mathbf{B} = \mu_0 \mathbf{j} + \frac{1}{c^2} \frac{\partial \mathbf{E}}{\partial t}, \qquad \nabla \cdot \mathbf{B} = 0, \qquad \nabla \times \mathbf{E} = -\frac{\partial \mathbf{B}}{\partial t}. \tag{1}$$

$$\partial_\beta \partial^\beta \varphi = \frac{1}{c^2} \frac{\partial^2 \varphi}{\partial t^2} - \Delta \varphi = \frac{\gamma \rho_{0q}}{\varepsilon_0}, \qquad \partial_\beta \partial^\beta \mathbf{A} = \frac{1}{c^2} \frac{\partial^2 \mathbf{A}}{\partial t^2} - \Delta \mathbf{A} = \mu_0 \mathbf{j}. \tag{2}$$

$$\mathbf{E} = -\nabla \varphi - \frac{\partial \mathbf{A}}{\partial t}, \qquad \mathbf{B} = \nabla \times \mathbf{A}, \qquad A_\mu = \left( \frac{\varphi}{c}, -\mathbf{A} \right). \tag{3}$$

For the particles moving inside the rotating sphere: $\gamma = \frac{1}{\sqrt{1 - v^2/c^2}}$ is the Lorentz factor; $\mathbf{v}$ is the particles' velocity in the reference frame $K$, in which the sphere is rotating; $\rho_{0q}$ is the charge density of a moving particle in the comoving reference frame; $\varepsilon_0$ is the electrical constant; $\mu_0$ is the magnetic constant; $\mathbf{j} = \gamma \rho_{0q} \mathbf{v}$ denotes the vector of the electric current density; $c$ is the speed of light, while $\mu_0 \varepsilon_0 c^2 = 1$; $A_\mu$ is the four-potential of the electromagnetic field; $\varphi$ and $\mathbf{A}$ are the scalar and vector potentials. Wave equations (2) for the potentials are obtained from equations (1) taking into account (3).

If the sphere with the particles rotates at a constant angular velocity $\omega$, the potentials would not depend on time. Then the time derivatives disappear in (2) and the following remains:

$$\Delta \varphi = -\frac{\gamma \rho_{0q}}{\varepsilon_0}, \qquad \Delta \mathbf{A} = -\mu_0 \mathbf{j} = -\mu_0 \gamma \rho_{0q} \mathbf{v}. \tag{4}$$

Equations (4) were solved in the absence of rotation, when $\omega = 0$, for a relativistic uniform system [11]. In this case, the Lorentz factor $\gamma'$ of the particles' motion relative to the reference frame $K'$, associated with the center of the fixed sphere, was substituted instead of $\gamma$ in (4). For the spherical system with the particles in the absence of the matter's general rotation the Lorentz factor according to [8] is equal to:

$$\gamma'(\omega=0)=\frac{c\gamma'_c}{r\sqrt{4\pi\eta\rho_0}}\sin\left(\frac{r}{c}\sqrt{4\pi\eta\rho_0}\right)\approx\gamma'_c-\frac{2\pi\eta\rho_0 r^2\gamma'_c}{3c^2}. \tag{5}$$

In (5) $r$ is the current radius, $\gamma'_c$ is the Lorentz factor at the center of the sphere, $\eta$ is the acceleration field coefficient, $\rho_0$ is the mass density of a moving particle in the comoving reference frame. Taking this into account, the scalar (electric) potential $\varphi_i$ inside the sphere and the similar potential $\varphi_o$ outside the sphere are defined by the expressions:

$$\varphi_i=\frac{\rho_{0q}c^2\gamma'_c}{4\pi\varepsilon_0\eta\rho_0 r}\left[\frac{c}{\sqrt{4\pi\eta\rho_0}}\sin\left(\frac{r}{c}\sqrt{4\pi\eta\rho_0}\right)-r\cos\left(\frac{a}{c}\sqrt{4\pi\eta\rho_0}\right)\right]\approx\frac{\rho_{0q}\gamma'_c(3a^2-r^2)}{6\varepsilon_0}. \tag{6}$$

$$\varphi_o=\frac{\rho_{0q}c^2\gamma'_c}{4\pi\varepsilon_0\eta\rho_0 r}\left[\frac{c}{\sqrt{4\pi\eta\rho_0}}\sin\left(\frac{a}{c}\sqrt{4\pi\eta\rho_0}\right)-a\cos\left(\frac{a}{c}\sqrt{4\pi\eta\rho_0}\right)\right].$$

$$\varphi_o=\frac{q_b}{4\pi\varepsilon_0 r}\approx\frac{q\gamma'_c}{4\pi\varepsilon_0 r}\left(1-\frac{3\eta m}{10ac^2}\right). \tag{7}$$

In (7) the quantity $q$ is the product of $\rho_{0q}$ by the volume $V_s$ of the sphere of radius $a$, that is $q=\frac{4\pi\rho_{0q}a^3}{3}$. Similarly, $m$ is the product of the invariant mass density $\rho_0$ of the matter's particles by the sphere's volume. However, the external potential $\varphi_o$ of the electric field does not depend on $q$, but depends on the total charge $q_b$ of the sphere, defined by the expression:

$$q_b = \rho_{0q} \int \gamma'(\omega = 0) dV_s = \frac{\rho_{0q} c^2 \gamma'_c}{\eta \rho_0} \left[ \frac{c}{\sqrt{4\pi \eta \rho_0}} \sin\left( \frac{a}{c} \sqrt{4\pi \eta \rho_0} \right) - a \cos\left( \frac{a}{c} \sqrt{4\pi \eta \rho_0} \right) \right] \approx$$

$$\approx \frac{4\pi \rho_{0q} a^3 \gamma'_c}{3} \left( 1 - \frac{2\pi \eta \rho_0 a^2}{5c^2} \right) = q\gamma'_c \left( 1 - \frac{3\eta m}{10ac^2} \right) \tag{8}$$

As for the vector (magnetic) potential $\mathbf{A}$ in (4), on the average it turns out to be equal to zero everywhere due to the chaotic motion of particles.

The particles' rotation at the angular velocity $\omega$ about the axis $OZ$ that passes through the center of the sphere changes the particles' linear velocities. Taking into account the rule of relativistic addition of velocities, for the absolute velocity and the Lorentz factor of an arbitrary particle we find the following:

$$\mathbf{v} = \frac{\mathbf{v}' + \frac{(\gamma_r - 1)(\mathbf{v}'\mathbf{v}_r)}{v_r^2} \mathbf{v}_r + \gamma_r \mathbf{v}_r}{\gamma_r \left( 1 + \frac{\mathbf{v}'\mathbf{v}_r}{c^2} \right)}, \qquad \gamma = \gamma' \gamma_r \left( 1 + \frac{\mathbf{v}'\mathbf{v}_r}{c^2} \right), \tag{9}$$

where $\mathbf{v}'$ is the velocity of chaotic motion of a particle in the reference frame $K'$ rotating with the matter at the angular velocity $\omega$; $\mathbf{v}_r$ is the linear velocity of motion of the reference frame $K'$ at the particle's location, arising due to rotation in the reference frame $K$; $\gamma_r = \frac{1}{\sqrt{1 - v_r^2/c^2}}$ is the Lorentz factor for the velocity $\mathbf{v}_r$, $\gamma' = \frac{1}{\sqrt{1 - v'^2/c^2}}$ is the Lorentz factor for the velocity $\mathbf{v}'$.

Expressions (9) should be averaged over the volume in a small neighborhood of the point under consideration so that a sufficient number of particles would be present in this volume. Due to the chaotic character of motion, the velocities $\mathbf{v}'$ of neighboring particles are directed in different ways. As a result, the average values will be: $\overline{\mathbf{v}} = \mathbf{v}_r$, $\overline{\gamma} = \gamma' \gamma_r$. Next, we will assume that, despite the general rotation, formula (5) for $\gamma'$ continues to be valid in the reference frame $K'$, with the exception that instead of the Lorentz factor $\gamma'_c$ at the center of the sphere, the formula must contain a quantity denoted as $\gamma_c$. Indeed, $\gamma'_c$ is determined in the absence of rotation, but the Lorentz factor at the center of the sphere can be changed due to rotation and turn into $\gamma_c$.

### 2.1. Potentials outside the rotating sphere

The charge density $\rho_{0q}$ outside the sphere is zero due to the absence of charged particles there. This simplifies the form of equations (4), and they turn into Laplace equations:

$$\Delta\varphi = 0, \qquad\qquad \Delta\mathbf{A} = 0. \tag{10}$$

From the great number of possible solutions of equations (10), we should choose those that, in the absence of rotation, go over to the solution of (7) for the scalar potential $\varphi_o$ and to the solution $\mathbf{A}_o = 0$ for the vector potential.

In order to find the necessary solutions, we will use the Lienard-Wiechert approach for retarded potentials. Let us assume that a point charged particle rotates along a circle of radius $\rho$ at the angular velocity $\omega$ and with the linear velocity $v_r = \omega\rho$. We will place the cylindrical reference frame with coordinates $\rho, \phi, z_d$ at the center of the sphere and will search for the electromagnetic field potentials from the rotating charge at a certain remote point $P$ with the radius vector $\mathbf{R} = (x, y, z)$.

The current position of the charge is given by the radius vector

$$\mathbf{r}_q = (\rho\cos\phi, \rho\sin\phi, z_d) = \left[\rho\cos\left(\omega t + \phi_0\right), \rho\sin\left(\omega t + \phi_0\right), z_d\right],$$

so that the circle of rotation is parallel to the plane $XOY$, while the angle $\phi$ depends on the current time: $\phi = \omega t + \phi_0$, here the constant $\phi_0$ is the initial phase.

The vector from the charge to the point $P$ will be as follows:

$$\mathbf{R}_P = \mathbf{R} - \mathbf{r}_q = \left[x - \rho\cos\left(\omega t + \phi_0\right), y - \rho\sin\left(\omega t + \phi_0\right), z - z_d\right],$$

wherein

$$R_P = \sqrt{\left(x-\rho\cos\phi\right)^2 + \left(y-\rho\sin\phi\right)^2 + \left(z-z_d\right)^2} = \\ = \sqrt{R^2 + z_d^2 - 2zz_d + \rho^2 - 2\rho x\cos\phi - 2\rho y\sin\phi}. \tag{11}$$

The Lienard-Wiechert formulas for the scalar and vector potentials of one particle with the number $n$ have the following form:

$$\varphi_n = \frac{q_n}{4\pi\varepsilon_0\left(\hat{R}_P - \hat{\mathbf{v}}\cdot\hat{\mathbf{R}}_P / c\right)}, \qquad \mathbf{A}_n = \frac{\mu_0 q_n \hat{\mathbf{v}}}{4\pi\left(\hat{R}_P - \hat{\mathbf{v}}\cdot\hat{\mathbf{R}}_P / c\right)}. \tag{12}$$

Here $\hat{\mathbf{R}}_P = \mathbf{R} - \hat{\mathbf{r}}_q$ is the vector from the charge to the point $P$ at the early time point $\hat{t} = t - \frac{\hat{R}_P}{c}$, the radius vector

$$\hat{\mathbf{r}}_q = (\rho\cos\hat{\phi}, \rho\sin\hat{\phi}, z_p) = \left[\rho\cos\left(\omega\hat{t}+\phi_0\right), \rho\sin\left(\omega\hat{t}+\phi_0\right), z_d\right]$$

defines the position of the charge at the time point $\hat{t}$, while

$$\hat{R}_P = \sqrt{\left(x-\rho\cos\hat{\phi}\right)^2 + \left(y-\rho\sin\hat{\phi}\right)^2 + \left(z-z_d\right)^2}.$$

The current rotation velocity of the charge is $\mathbf{v}_r = \frac{d\mathbf{r}_q}{dt} = (-\omega\rho\sin\phi, \omega\rho\cos\phi, 0)$, and the charge's velocity at the early time point will be $\hat{\mathbf{v}}_r = \mathbf{v}_r(\hat{t}) = \frac{d\hat{\mathbf{r}}_q}{d\hat{t}} = (-\omega\rho\sin\hat{\phi}, \omega\rho\cos\hat{\phi}, 0)$, wherein $\hat{\phi} = \omega\hat{t} + \phi_0 = \omega t + \phi_0 - \frac{\omega\hat{R}_P}{c} = \phi - \frac{\omega\hat{R}_P}{c} = \phi - \phi_P$.

Since, according to (9), the average velocity of the particles' motion is $\overline{\mathbf{v}} = \mathbf{v}_r$, in (12) $\hat{\mathbf{v}}_r$ should be used instead of $\hat{\mathbf{v}}$. Then for $\hat{R}_P$ and the product $\hat{\mathbf{v}}_r \cdot \hat{\mathbf{R}}_P$ in (12) we obtain:

$$\hat{R}_P = \sqrt{R^2 + z_d^2 - 2zz_d + \rho^2 - 2\rho x\cos\hat{\phi} - 2\rho y\sin\hat{\phi}}, \qquad \hat{\mathbf{v}}_r \cdot \hat{\mathbf{R}}_P = \omega\rho y\cos\hat{\phi} - \omega\rho x\sin\hat{\phi}.$$

(13)

Let’s locate the coordinate $z_d$ in such a way that it would define the location of a thin layer with thickness $s$ in the form of a disk parallel to the plane $XOY$. The radius of such a disk inside the sphere will be $\rho_d = \sqrt{a^2 - z_d^2}$, where the sphere’s radius is $a$. The sphere is tightly filled with rotating particles, and the same applies to this disk. We will use the principle of superposition of potentials and will find the scalar potential at the remote point $P$ from the rotating disk with charged particles. For this purpose we need to take the sum over all $N$ charges in the disk. In view of (12), for the scalar potential we have the following:

$$\varphi_d = \sum_{n=1}^{N} \varphi_n = \frac{1}{4\pi\varepsilon_0} \sum_{n=1}^{N} \frac{q_n}{\left(\hat{R}_P - \hat{\mathbf{v}}_r \cdot \hat{\mathbf{R}}_P / c\right)_n}. \quad (14)$$

Each charge $q_n$ inside the disk has its own rotation radius $\rho_n$, motion velocity $v_n = \omega \rho_n$, while the instantaneous position of the charge is given by the vector $\mathbf{r}_{qn} = (\rho_n \cos\phi_n, \rho_n \sin\phi_n, z_d)$. In this regard, in (14) the denominator depends on the location of the particle in the disk, and therefore it has an index $n$.

The charge of a point particle rotating in the disk can be expressed in terms of the invariant charge density, Lorentz factor, and moving volume:

$$q_n = \frac{\rho_{0q} \gamma s \rho \, d\rho \, d\phi}{\gamma_r}.$$

The quantity $\frac{s\rho \, d\rho \, d\phi}{\gamma_r}$ here specifies the element of the volume of a rotating disk, which, as a result of Lorentz contraction, is $\gamma_r$ times less than the volume element $s\rho d\rho d\phi$ of the fixed disk. The quantity $\rho_{0q}\gamma$ defines the effective density of the charge, taking into account its rotation inside the disk and the chaotic motion of particles. As $\gamma$ in (15) we should substitute the averaged value of the Lorentz factor $\bar{\gamma} = \gamma' \gamma_r$, according to (9). This gives the following:

$$q_n = \rho_{0q} \gamma' s \rho \, d\rho \, d\phi. \quad (15)$$

The charge $q_n$ is expressed in terms of the product of differentials, so that the sum (14) can be transformed into an integral. With this in mind, from (13-15) it follows:

$$\varphi_d = \frac{s\rho_{0q}}{4\pi\varepsilon_0}\int_0^{2\pi}\int_0^{\rho_d}\frac{\gamma'\rho d\rho d\phi}{\sqrt{R^2+z_d^2-2zz_d+\rho^2-2\rho x\cos\hat{\phi}-2\rho y\sin\hat{\phi}}+\frac{\omega\rho x\sin\hat{\phi}}{c}-\frac{\omega\rho y\cos\hat{\phi}}{c}}. \tag{16}$$

In order to be able to perform integration, in (16) we need to express the angle $\hat{\phi}$, defining the position of an arbitrary particle at the early time point $\hat{t}$, in terms of the angle $\phi$ at the time point $t$. Since $\phi = \omega t + \phi_0$, $\hat{\phi} = \omega\hat{t} + \phi_0$, $\hat{t} = t - \frac{\hat{R}_P}{c}$, then we will have $\hat{\phi} = \phi - \frac{\omega\hat{R}_P}{c}$, and therefore

$$\cos\hat{\phi} = \cos\phi\cos\frac{\omega\hat{R}_P}{c} + \sin\phi\sin\frac{\omega\hat{R}_P}{c}, \qquad \sin\hat{\phi} = \sin\phi\cos\frac{\omega\hat{R}_P}{c} - \cos\phi\sin\frac{\omega\hat{R}_P}{c}. \tag{17}$$

From comparison of (12) and (16) it follows that the vector potential of the rotating disk will be equal to:

$$\mathbf{A}_d = \frac{\mu_0 s\rho_{0q}}{4\pi}\int_0^{2\pi}\int_0^{\rho_d}\frac{\gamma'\,\hat{\mathbf{v}}_r\,\rho d\rho d\phi}{\sqrt{R^2+z_d^2-2zz_d+\rho^2-2\rho x\cos\hat{\phi}-2\rho y\sin\hat{\phi}}+\frac{\omega\rho x\sin\hat{\phi}}{c}-\frac{\omega\rho y\cos\hat{\phi}}{c}}.$$

In (16), the scalar potential $\varphi_d$ is sought for the remote point $P$ with the radius vector $\mathbf{R}=(x,y,z)$. The vector potential $\mathbf{A}_d$ at this point depends on the velocity $\hat{\mathbf{v}}_r = \mathbf{v}_r(\hat{t}) = (-\omega\rho\sin\hat{\phi}, \omega\rho\cos\hat{\phi}, 0)$ of motion of the charged particles of the rotating disk at the early time $\hat{t}$. The velocity $\hat{\mathbf{v}}_r$ lies in a plane parallel to the plane $XOY$, and the same holds true for $\mathbf{A}_d$. For the components $\mathbf{A}_d$, we can write the following:

$$A_{dx} = -\frac{\mu_0 \omega s \rho_{0q}}{4\pi} \int_0^{2\pi} \int_0^{\rho_d} \frac{\gamma' \sin\hat{\phi}\, \rho^2 d\rho d\phi}{\sqrt{R^2 + z_d^2 - 2zz_d + \rho^2 - 2\rho x \cos\hat{\phi} - 2\rho y \sin\hat{\phi}} + \frac{\omega \rho x \sin\hat{\phi}}{c} - \frac{\omega \rho y \cos\hat{\phi}}{c}} .$$

$$A_{dy} = \frac{\mu_0 \omega s \rho_{0q}}{4\pi} \int_0^{2\pi} \int_0^{\rho_d} \frac{\gamma' \cos\hat{\phi}\, \rho^2 d\rho d\phi}{\sqrt{R^2 + z_d^2 - 2zz_d + \rho^2 - 2\rho x \cos\hat{\phi} - 2\rho y \sin\hat{\phi}} + \frac{\omega \rho x \sin\hat{\phi}}{c} - \frac{\omega \rho y \cos\hat{\phi}}{c}} .$$

$$A_{dz} = 0 . \tag{18}$$

### 2.2. Scalar potential in the middle zone

Let us first consider the case when in (17) the conditions $\hat{R}_P >> a$, $\frac{\omega \hat{R}_P}{c} << 1$ are met, which corresponds to the case of sufficiently large distances $R$ from the sphere of radius $a$ to the point $P$ where the scalar potential is sought. As an example, let us assume that the relations of sizes and velocities are given by the relative value of 1%. In this case, the condition of the middle zone at $R \approx \hat{R}_P$ means that it should be $\frac{a}{R} < 0,01$ and $\frac{\omega R}{c} < 0,01$, so that a two-sided inequality $100a < R < \frac{c}{100\omega}$ is obtained for the distance.

Under the above conditions for $\hat{R}_P$ we can assume in (17) that

$$\cos\hat{\phi} \approx \cos\phi + \frac{\omega \hat{R}_P}{c} \sin\phi , \qquad \sin\hat{\phi} \approx \sin\phi - \frac{\omega \hat{R}_P}{c} \cos\phi . \tag{19}$$

Let us square $\hat{R}_P$ in (13), substitute there $\cos\hat{\phi}$ and $\sin\hat{\phi}$ from (19), obtain a quadratic equation to determine $\hat{R}_P$ and write down its solution:

$$\hat{R}_P^2 + \frac{2\omega\rho\left(x\sin\phi - y\cos\phi\right)}{c} \hat{R}_P - R^2 - z_d^2 + 2zz_d - \rho^2 + 2\rho x \cos\phi + 2\rho y \sin\phi = 0 .$$

$$\hat{R}_P=-\frac{\omega\rho\left(x\sin\phi-y\cos\phi\right)}{c}+$$
$$+\sqrt{R^2+z_d^2-2zz_d+\rho^2-2\rho x\cos\phi-2\rho y\sin\phi+\frac{\omega^2\rho^2\left(x\sin\phi-y\cos\phi\right)^2}{c^2}}. \tag{20}$$

Since the square root in (16) is equal to $\hat{R}_P$ according to (13), then we can replace this square root with the expression for $\hat{R}_P$ from (20). Then using $\sin\hat{\phi}$ and $\cos\hat{\phi}$ from (19) for transformation of (16), we arrive at the expression:

$$\varphi_d=\frac{s\rho_{0q}}{4\pi\varepsilon_0}\int_0^{2\pi}\int_0^{\rho_d}\frac{\gamma'\rho\,d\rho\,d\phi}{\sqrt{\begin{array}{l}R^2+z_d^2-2zz_d+\rho^2-2\rho x\cos\phi-\\-2\rho y\sin\phi+\frac{\omega^2\rho^2\left(x\sin\phi-y\cos\phi\right)^2}{c^2}\end{array}}-\frac{\omega^2\rho x\hat{R}_P}{c^2}\cos\phi-\frac{\omega^2\rho y\hat{R}_P}{c^2}\sin\phi}. \tag{21}$$

In (21) we will expand the square root to the third-order terms by the rule $\sqrt{1+\delta}\approx1+\frac{\delta}{2}-\frac{\delta^2}{8}$:

$$\sqrt{R^2+z_d^2-2zz_d+\rho^2-2\rho x\cos\phi-2\rho y\sin\phi+\frac{\omega^2\rho^2\left(x\sin\phi-y\cos\phi\right)^2}{c^2}}\approx$$
$$\approx\sqrt{R^2+z_d^2-2zz_d+\rho^2}\left[\begin{array}{l}1-\frac{\rho x\cos\phi+\rho y\sin\phi}{R^2+z_d^2-2zz_d+\rho^2}+\\+\frac{\omega^2\rho^2\left(x\sin\phi-y\cos\phi\right)^2}{2c^2\left(R^2+z_d^2-2zz_d+\rho^2\right)}-\frac{\rho^2\left(x\cos\phi+y\sin\phi\right)^2}{2\left(R^2+z_d^2-2zz_d+\rho^2\right)^2}+\\+\frac{\omega^2\rho^3\left(x\sin\phi-y\cos\phi\right)^2\left(x\cos\phi+y\sin\phi\right)}{2c^2\left(R^2+z_d^2-2zz_d+\rho^2\right)^2}\end{array}\right]. \tag{22}$$

Let us substitute (22) into (20):

$$\hat{R}_P \approx \sqrt{R^2+z_d^2-2zz_d+\rho^2}\begin{bmatrix} 1-\dfrac{\omega\rho\left(x\sin\phi-y\cos\phi\right)}{c\sqrt{R^2+z_d^2-2zz_d+\rho^2}}-\dfrac{\rho x\cos\phi+\rho y\sin\phi}{R^2+z_d^2-2zz_d+\rho^2}+ \\ +\dfrac{\omega^2\rho^2\left(x\sin\phi-y\cos\phi\right)^2}{2c^2\left(R^2+z_d^2-2zz_d+\rho^2\right)}-\dfrac{\rho^2\left(x\cos\phi+y\sin\phi\right)^2}{2\left(R^2+z_d^2-2zz_d+\rho^2\right)^2}+ \\ +\dfrac{\omega^2\rho^3\left(x\sin\phi-y\cos\phi\right)^2\left(x\cos\phi+y\sin\phi\right)}{2c^2\left(R^2+z_d^2-2zz_d+\rho^2\right)^2} \end{bmatrix}. \tag{23}$$

With the help of $\hat{R}_P$ from (23), we will transform the second and third terms in the denominator of (21), leaving only the terms containing $c^2$ and $c^3$:

$$-\frac{\omega^2\rho x\hat{R}_P}{c^2}\cos\phi-\frac{\omega^2\rho y\hat{R}_P}{c^2}\sin\phi\approx$$

$$\approx -\sqrt{R^2+z_d^2-2zz_d+\rho^2}\begin{bmatrix} \dfrac{\omega^2\rho\left(x\cos\phi+y\sin\phi\right)}{c^2}- \\ -\dfrac{\omega^3\rho^2\left(x\sin\phi-y\cos\phi\right)\left(x\cos\phi+y\sin\phi\right)}{c^3\sqrt{R^2+z_d^2-2zz_d+\rho^2}}- \\ -\dfrac{\omega^2\rho^2\left(x\cos\phi+y\sin\phi\right)^2}{c^2\left(R^2+z_d^2-2zz_d+\rho^2\right)}-\dfrac{\omega^2\rho^3\left(x\cos\phi+y\sin\phi\right)^3}{2c^2\left(R^2+z_d^2-2zz_d+\rho^2\right)^2} \end{bmatrix}. \tag{24}$$

Let us now substitute (22) and (24) into (21) and put $\sqrt{R^2+z_d^2-2zz_d+\rho^2}$ outside the brackets:

$$\varphi_d \approx \frac{s\rho_{0q}}{4\pi\varepsilon_0}\int_0^{2\pi}\int_0^{\rho_d} \frac{1}{\left[\begin{array}{l} 1-\dfrac{\rho\left(x\cos\phi+y\sin\phi\right)}{R^2+z_d^2-2zz_d+\rho^2}-\dfrac{\rho^2\left(x\cos\phi+y\sin\phi\right)^2}{2\left(R^2+z_d^2-2zz_d+\rho^2\right)^2}- \\ -\dfrac{\omega^2\rho\left(x\cos\phi+y\sin\phi\right)}{c^2}+ \\ +\dfrac{\omega^2\rho^2\left(x\sin\phi-y\cos\phi\right)^2}{2c^2\left(R^2+z_d^2-2zz_d+\rho^2\right)}+\dfrac{\omega^2\rho^2\left(x\cos\phi+y\sin\phi\right)^2}{c^2\left(R^2+z_d^2-2zz_d+\rho^2\right)}+ \\ +\dfrac{\omega^2\rho^3\left(x\cos\phi+y\sin\phi\right)\left(x\sin\phi-y\cos\phi\right)^2}{2c^2\left(R^2+z_d^2-2zz_d+\rho^2\right)^2}+ \\ +\dfrac{\omega^2\rho^3\left(x\cos\phi+y\sin\phi\right)^3}{2c^2\left(R^2+z_d^2-2zz_d+\rho^2\right)^2}+ \\ +\dfrac{\omega^3\rho^2\left(x\sin\phi-y\cos\phi\right)\left(x\cos\phi+y\sin\phi\right)}{c^3\sqrt{R^2+z_d^2-2zz_d+\rho^2}} \end{array}\right]} \frac{\gamma'\rho\, d\rho\, d\phi}{\sqrt{R^2+z_d^2-2zz_d+\rho^2}}.$$

In this integral, we will use an approximate expression of the form $\frac{1}{1+\delta}\approx 1-\delta+\delta^2$ for small $\delta$. This gives the following:

$$\varphi_d \approx \frac{s\rho_{0q}}{4\pi\varepsilon_0}\int_0^{2\pi}\int_0^{\rho_d} \frac{D\gamma'\rho\, d\rho\, d\phi}{\sqrt{R^2+z_d^2-2zz_d+\rho^2}} \,. \qquad (25)$$

The quantity $D$ in (25) is given by the expression:

$$D \approx 1+\frac{\rho x\cos\phi+\rho y\sin\phi}{R^2+z_d^2-2zz_d+\rho^2}+\frac{3\rho^2\left(x\cos\phi+y\sin\phi\right)^2}{2\left(R^2+z_d^2-2zz_d+\rho^2\right)^2}+\frac{\rho^3\left(x\cos\phi+y\sin\phi\right)^3}{\left(R^2+z_d^2-2zz_d+\rho^2\right)^3}+$$
$$+\frac{\rho^4\left(x\cos\phi+y\sin\phi\right)^4}{4\left(R^2+z_d^2-2zz_d+\rho^2\right)^4}+\frac{\omega^2\rho\left(x\cos\phi+y\sin\phi\right)}{c^2}-\frac{\omega^2\rho^2\left(x^2+y^2\right)}{2c^2\left(R^2+z_d^2-2zz_d+\rho^2\right)}+$$
$$+\frac{3\omega^2\rho^2\left(x\cos\phi+y\sin\phi\right)^2}{2c^2\left(R^2+z_d^2-2zz_d+\rho^2\right)}-\frac{3\omega^2\rho^3\left(x^2+y^2\right)\left(x\cos\phi+y\sin\phi\right)}{2c^2\left(R^2+z_d^2-2zz_d+\rho^2\right)^2}-$$
$$-\frac{2\omega^2\rho^4\left(x^2+y^2\right)\left(x\cos\phi+y\sin\phi\right)^2}{c^2\left(R^2+z_d^2-2zz_d+\rho^2\right)^3}+\frac{\omega^2\rho^4\left(x\cos\phi+y\sin\phi\right)^2\left(x\sin\phi-y\cos\phi\right)^2}{2c^2\left(R^2+z_d^2-2zz_d+\rho^2\right)^3}-$$
$$-\frac{\omega^2\rho^5\left(x^2+y^2\right)\left(x\cos\phi+y\sin\phi\right)^3}{2c^2\left(R^2+z_d^2-2zz_d+\rho^2\right)^4}-\frac{\omega^3\rho^2\left(x\sin\phi-y\cos\phi\right)\left(x\cos\phi+y\sin\phi\right)}{c^3\sqrt{R^2+z_d^2-2zz_d+\rho^2}}-$$
$$-\frac{2\omega^3\rho^3\left(x\sin\phi-y\cos\phi\right)\left(x\cos\phi+y\sin\phi\right)^2}{c^3\left(R^2+z_d^2-2zz_d+\rho^2\right)^{3/2}}-\frac{\omega^3\rho^4\left(x\sin\phi-y\cos\phi\right)\left(x\cos\phi+y\sin\phi\right)^3}{c^3\left(R^2+z_d^2-2zz_d+\rho^2\right)^{5/2}}.$$

(26)

In (25), only the quantity $D$ depends on the angle $\phi$, according to (26). After integration over this angle in (25), the following remains:

$$\varphi_d \approx \frac{s\rho_{0q}}{2\varepsilon_0}\int_0^{\rho_d}\frac{\left[\begin{array}{l}1+\dfrac{\omega^2\rho^2\left(x^2+y^2\right)}{4c^2\left(R^2+z_d^2-2zz_d+\rho^2\right)}+\dfrac{3\rho^2\left(x^2+y^2\right)}{4\left(R^2+z_d^2-2zz_d+\rho^2\right)^2}-\\-\dfrac{15\omega^2\rho^4\left(x^2+y^2\right)^2}{16c^2\left(R^2+z_d^2-2zz_d+\rho^2\right)^3}+\dfrac{3\rho^4\left(x^2+y^2\right)^2}{32\left(R^2+z_d^2-2zz_d+\rho^2\right)^4}\end{array}\right]\gamma'\rho\,d\rho}{\sqrt{R^2+z_d^2-2zz_d+\rho^2}}.$$

The last two terms in the square brackets inside the integral, due to their smallness, can be further neglected. The Lorentz factor $\gamma'$, similarly to (5), is written in the first approximation as follows:

$$\gamma'=\frac{c\gamma_c}{r\sqrt{4\pi\eta\rho_0}}\sin\left(\frac{r}{c}\sqrt{4\pi\eta\rho_0}\right)\approx\gamma_c-\frac{2\pi\eta\rho_0 r^2\gamma_c}{3c^2}=\gamma_c-\frac{2\pi\eta\rho_0\left(\rho^2+z_d^2\right)\gamma_c}{3c^2},$$

(27)

We will note that here we use the Lorentz factor $\gamma_c$ at the center of the rotating sphere, which may not be equal to $\gamma'_c$ in (5) at the center of the sphere at rest. In view of $\gamma'$ from (27), we can write for $\varphi_d$ the following:

$$\varphi_d \approx \frac{s\rho_{0q}\gamma_c}{2\varepsilon_0}\int_0^{\rho_d}\frac{\left[\begin{array}{l}1-\dfrac{2\pi\eta\rho_0\left(\rho^2+z_d^2\right)}{3c^2}+\\ +\dfrac{\omega^2\rho^2\left(x^2+y^2\right)}{4c^2\left(R^2+z_d^2-2zz_d+\rho^2\right)}+\dfrac{3\rho^2\left(x^2+y^2\right)}{4\left(R^2+z_d^2-2zz_d+\rho^2\right)^2}\end{array}\right]\rho\,d\rho}{\sqrt{R^2+z_d^2-2zz_d+\rho^2}}.$$

We will represent the potential as the sum of four terms, obtained by integrating the potential $\varphi_d$ over the variable $\rho$:

$$\varphi_d \approx \frac{s\rho_{0q}\gamma_c}{2\varepsilon_0}\left(I_1+I_2+I_3+I_4\right),$$

where

$$I_1=\int_0^{\rho_d}\frac{\rho\,d\rho}{\sqrt{R^2+z_d^2-2zz_d+\rho^2}}, \qquad I_2=-\frac{2\pi\eta\rho_0}{3c^2}\int_0^{\rho_d}\frac{\left(\rho^2+z_d^2\right)\rho\,d\rho}{\sqrt{R^2+z_d^2-2zz_d+\rho^2}},$$

$$I_3=\frac{\omega^2\left(x^2+y^2\right)}{4c^2}\int_0^{\rho_d}\frac{\rho^3\,d\rho}{\left(R^2+z_d^2-2zz_d+\rho^2\right)^{3/2}}, \qquad I_4=\frac{3\left(x^2+y^2\right)}{4}\int_0^{\rho_d}\frac{\rho^3\,d\rho}{\left(R^2+z_d^2-2zz_d+\rho^2\right)^{5/2}}.$$

These integrals, taking into account the relation $\rho_d^2=a^2-z_d^2$, equal to:

$$I_1=\sqrt{R^2+a^2-2zz_d}-\sqrt{R^2+z_d^2-2zz_d}.$$

$$I_2=-\frac{2\pi\eta\rho_0}{3c^2}\left[\frac{\left(R^2+a^2-2zz_d\right)^{3/2}}{3}-\left(R^2+z_d^2-2zz_d\right)\sqrt{R^2+a^2-2zz_d}+ \atop +\frac{2\left(R^2+z_d^2-2zz_d\right)^{3/2}}{3}\right]-$$

$$-\frac{2\pi\eta\rho_0 z_d^2}{3c^2}\left(\sqrt{R^2+a^2-2zz_d}-\sqrt{R^2+z_d^2-2zz_d}\right).$$

$$I_3=\frac{\omega^2\left(x^2+y^2\right)}{4c^2}\left[\sqrt{R^2+a^2-2zz_d}+\frac{R^2+z_d^2-2zz_d}{\sqrt{R^2+a^2-2zz_d}}-2\sqrt{R^2+z_d^2-2zz_d}\right].$$

$$I_4=-\frac{\left(x^2+y^2\right)}{4}\left[\frac{a^2-z_d^2}{\left(R^2+a^2-2zz_d\right)^{3/2}}+\frac{2}{\sqrt{R^2+a^2-2zz_d}}-\frac{2}{\sqrt{R^2+z_d^2-2zz_d}}\right].$$

(28)

The potential $\varphi_d$ is the potential at the remote point $P$ from one thin layer in the form of a disk, which is parallel to the plane $XOY$ and shifted along the axis $OZ$ by distance $z_d$. Now it is necessary to sum up separate potentials created at the point $P$ by all layers of the ball, taking into account that the layer thickness $s$ is the differential $dz_d$. Passing from the sum to the integral, we find:

$$\varphi\approx\frac{\rho_{0q}\gamma_c}{2\varepsilon_0}\int_{-a}^{a}\left(I_1+I_2+I_3+I_4\right)dz_d\,.$$

Using (28), we have the following:

$$\int_{-a}^{a}I_1\,dz_d\approx\frac{2a^3}{3R}, \qquad \int_{-a}^{a}I_2\,dz_d\approx-\frac{4\pi\eta\rho_0 a^5}{15c^2R},$$

$$\int_{-a}^{a}I_3\,dz_d\approx\frac{\omega^2a^5\left(x^2+y^2\right)}{15c^2R^3}, \qquad \int_{-a}^{a}I_4\,dz_d\approx\frac{a^5\left(x^2+y^2\right)}{5R^5}.$$

Taking this into account, the following is obtained for the potential:

$$\varphi \approx \frac{\rho_{0q} a^3 \gamma_c}{3\varepsilon_0 R} \left[ 1 - \frac{2\pi \eta \rho_0 a^2}{5c^2} + \frac{3a^2 \left( x^2 + y^2 \right)}{10R^4} + \frac{\omega^2 a^2 \left( x^2 + y^2 \right)}{10c^2 R^2} \right]. \tag{29}$$

Expression (29) for the potential is an approximate solution in the middle zone, where the conditions $R >> a$, $\frac{\omega \hat{R}_P}{c} \approx \frac{\omega R_P}{c} \approx \frac{\omega R}{c} << 1$ are met.

Let us now calculate the charge of a slowly rotating sphere in spherical coordinates $r, \theta, \phi$. According to (9), the averaged Lorentz factor of the particles' motion is $\bar{\gamma} = \gamma' \gamma_r$, the charge density inside the sphere is $\gamma' \gamma_r \rho_{0q}$, and the element of volume moving due to rotation is $dV_s = \frac{r^2 dr d\phi \sin\theta d\theta}{\gamma_r}$. Hence, in view of (27), we find for the charge by integrating over the sphere's volume in spherical coordinates:

$$q_\omega = \rho_{0q} \int \gamma' \gamma_r dV_s = \frac{c \rho_{0q} \gamma_c}{\sqrt{4\pi \eta \rho_0}} \int \sin\left( \frac{r}{c} \sqrt{4\pi \eta \rho_0} \right) r dr d\phi \sin\theta d\theta . \tag{30}$$

The integration result is as follows:

$$\begin{aligned} q_\omega &= \frac{\rho_{0q} c^2 \gamma_c}{\eta \rho_0} \left[ \frac{c}{\sqrt{4\pi \eta \rho_0}} \sin\left( \frac{a}{c} \sqrt{4\pi \eta \rho_0} \right) - a \cos\left( \frac{a}{c} \sqrt{4\pi \eta \rho_0} \right) \right] \approx \\ &\approx \frac{4\pi \rho_{0q} a^3 \gamma_c}{3} \left( 1 - \frac{2\pi \eta \rho_0 a^2}{5c^2} \right). \end{aligned} \tag{31}$$

According to the method of calculation in (31), the charge $q_\omega$ is the sum of invariant charges of all the system's particles and, therefore, is an invariant quantity that does not depend on the angular velocity of rotation $\omega$. In this case, the charge $q_\omega$ (31) must be equal to the charge $q_b$ in (8). Hence we find the equality of the Lorentz factors at the center of the sphere for the cases of a sphere at rest and a similar rotating sphere: $\gamma_c = \gamma'_c$.

From (29) and (31) it follows:

$$\varphi \approx \frac{q_\omega}{4\pi\varepsilon_0 R}\left[1+\frac{3a^2\left(x^2+y^2\right)}{10R^4}+\frac{\omega^2 a^2\left(x^2+y^2\right)}{10c^2R^2}\right]. \qquad (32)$$

In order to check the solution of (32) for the potential, we can substitute it in (10) into the equation $\Delta\varphi = \frac{\partial^2\varphi}{\partial x^2}+\frac{\partial^2\varphi}{\partial y^2}+\frac{\partial^2\varphi}{\partial z^2}=0$. As a consequence, it appears that in (32) the sum of the two terms $\frac{3a^2\left(x^2+y^2\right)}{10R^4}+\frac{\omega^2 a^2\left(x^2+y^2\right)}{10c^2R^2}$ does not agree with this equation. This is possible because during integration we neglected all the possible small terms, the presence of which could lead to satisfying the Laplace equation $\Delta\varphi = 0$. In this regard, we will remind that in (17) we expanded the sine and cosine only to the first-order terms in the form $\sin\left(\omega\hat{R}_P / c\right)\approx \omega\hat{R}_P / c$ , $\cos\left(\omega\hat{R}_P / c\right)\approx 1$, obtaining (19). But in (32) the angular velocity $\omega$ is present in the second-order term containing the square of the speed of light in the denominator. This term can change if in (17) we expand the sine and cosine to the second-order terms in the form $\sin\left(\frac{\omega\hat{R}_P}{c}\right)\approx \frac{\omega\hat{R}_P}{c}-\frac{\omega^3\hat{R}_P^3}{6c^3}$, $\cos\left(\frac{\omega\hat{R}_P}{c}\right)\approx 1-\frac{\omega^2\hat{R}_P^2}{2c^2}$. On the other hand, the presence of the small term $\frac{3a^2\left(x^2+y^2\right)}{10R^4}$ contradicts the Coulomb law at $\omega = 0$, and its very appearance may be a consequence of the adopted approximation procedure.

For the Laplace equation to hold, we will substitute the sum of the two terms in (32) with $\frac{\omega^2 a^2}{10c^2}$. At least such substitution is quite acceptable under the conditions $R >> a$, $x^2+y^2 >> z^2$, $x^2+y^2\approx R^2$. In view of the above, the potential takes the following form:

$$\varphi \approx \frac{q_\omega}{4\pi\varepsilon_0 R}\left(1+\frac{\omega^2 a^2}{10c^2}\right). \qquad (33)$$

Since $\frac{\omega R}{c} << 1, a << R$, $\frac{\omega a}{c} << 1$, the term in the brackets in (33) makes a very small correction to the potential $\varphi$. If we again assume that the relations of sizes and velocities in the system are given by the relative value of about 1%, then $\frac{a}{R} < 0{,}01$, $\frac{\omega R}{c} < 0{,}01$ and $\frac{\omega a}{c} < 0,0001$. In this case, the entire expression in brackets in (33) can make a correction to the potential not exceeding 0.0000001%.

It is easy to verify that potential (33) satisfies the Laplace equation $\Delta\varphi = 0$. In this case, for the sphere at rest the potential, according to (7), has the form $\varphi_o = \frac{q_b}{4\pi\varepsilon_0 R}$, while $q_b = q_\omega$, $\gamma'_c = \gamma_c$, and the equation $\Delta\varphi_o = 0$ holds.

It follows from the above that in the general case the potential outside the rotating sphere can be represented by the formula

$$\varphi = \frac{q_\omega}{4\pi\varepsilon_0 R} F, \tag{34}$$

where the function $F$ can be a function of $\omega$, $a$, $R$ and of $x^2 + y^2$. At $\omega = 0$ it must be $F = 1$, and in case of rotation of the sphere with charged particles, for the middle zone, where the conditions $R >> a$, $\frac{\omega R}{c} << 1$, $x^2 + y^2 \approx R^2$ are met, we must have $F \approx 1 + \frac{\omega^2 a^2}{10c^2}$ if we consider (33) true. Thus, the function $F$ differs little from 1.

It follows from expression (32) that in the middle zone the potential actually can depend on the direction to the observation point and, at the same distance $R$, it increases as it gets closer to the equatorial plane. This could be a consequence of the spherical-cylindrical symmetry of arrangement of the moving charges when the potential is calculated. Indeed, according to (16), the potential $\varphi_d$ from one disk inside the sphere depends on the retarded angle $\hat{\phi} = \omega\hat{t} + \phi_0 = \omega t + \phi_0 - \frac{\omega\hat{R}_P}{c} = \phi - \frac{\omega\hat{R}_P}{c}$, which is a function of the angular velocity $\omega$. This implies the dependence of the potential $\varphi$ outside the sphere on $\omega$, which can be realized in the form of (32-34).

### 2.3. Vector potential in the middle zone

Proceeding in the same way as when we obtained (25) from (16), we will transform the components of the vector potential of the rotating disk (18):

$$A_{dx} = -\frac{\mu_0 \omega s \rho_{0q}}{4\pi} \int_0^{2\pi} \int_0^{\rho_d} \frac{D\gamma' \sin\hat{\phi}\, \rho^2 \, d\rho \, d\phi}{\sqrt{R^2 + z_d^2 - 2zz_d + \rho^2}} \,,$$

$$A_{dy} \approx \frac{\mu_0 \omega s \rho_{0q}}{4\pi} \int_0^{2\pi} \int_0^{\rho_d} \frac{D\gamma' \cos\hat{\phi}\, \rho^2 \, d\rho \, d\phi}{\sqrt{R^2 + z_d^2 - 2zz_d + \rho^2}} \,, \qquad A_{dz} = 0 \,. \tag{35}$$

Substitution of $\hat{R}_p$ from (23) into (19) gives the following:

$$\cos\hat{\phi} \approx \cos\phi + \frac{\omega}{c}\sqrt{R^2 + z_d^2 - 2zz_d + \rho^2}\, \sin\phi - \frac{\omega^2 \rho \left( x\sin\phi - y\cos\phi \right) \sin\phi}{c^2} -$$
$$- \frac{\omega\rho\left( x\cos\phi + y\sin\phi \right)\sin\phi}{c\sqrt{R^2 + z_d^2 - 2zz_d + \rho^2}} + \frac{\omega^3 \rho^2 \left( x\sin\phi - y\cos\phi \right)^2 \sin\phi}{2c^3 \sqrt{R^2 + z_d^2 - 2zz_d + \rho^2}} -$$
$$- \frac{\omega\rho^2 \left( x\cos\phi + y\sin\phi \right)^2 \sin\phi}{2c\left( R^2 + z_d^2 - 2zz_d + \rho^2 \right)^{3/2}} + \frac{\omega^3 \rho^3 \left( x\sin\phi - y\cos\phi \right)^2 \left( x\cos\phi + y\sin\phi \right)\sin\phi}{2c^3 \left( R^2 + z_d^2 - 2zz_d + \rho^2 \right)^{3/2}} .$$

$$\sin\hat{\phi} \approx \sin\phi - \frac{\omega}{c}\sqrt{R^2 + z_d^2 - 2zz_d + \rho^2}\, \cos\phi + \frac{\omega^2 \rho \left( x\sin\phi - y\cos\phi \right) \cos\phi}{c^2} +$$
$$+ \frac{\omega\rho\left( x\cos\phi + y\sin\phi \right)\cos\phi}{c\sqrt{R^2 + z_d^2 - 2zz_d + \rho^2}} - \frac{\omega^3 \rho^2 \left( x\sin\phi - y\cos\phi \right)^2 \cos\phi}{2c^3 \sqrt{R^2 + z_d^2 - 2zz_d + \rho^2}} +$$
$$+ \frac{\omega\rho^2 \left( x\cos\phi + y\sin\phi \right)^2 \cos\phi}{2c\left( R^2 + z_d^2 - 2zz_d + \rho^2 \right)^{3/2}} - \frac{\omega^3 \rho^3 \left( x\sin\phi - y\cos\phi \right)^2 \left( x\cos\phi + y\sin\phi \right)\cos\phi}{2c^3 \left( R^2 + z_d^2 - 2zz_d + \rho^2 \right)^{3/2}} .$$

(36)

We will use $D$ from (26), as well as $\cos\hat{\phi}$ and $\sin\hat{\phi}$ from (36), and will integrate the products of these quantities over the angle $\phi$:

$$\int_0^{2\pi} D\sin\hat{\phi}\, d\phi \approx \frac{\rho y \pi}{R^2 + z_d^2 - 2zz_d + \rho^2} + \frac{3\rho^3 y\left( x^2 + y^2 \right)\pi}{4\left( R^2 + z_d^2 - 2zz_d + \rho^2 \right)^3} + \frac{3\omega\rho^3 x\left( x^2 + y^2 \right)\pi}{4c\left( R^2 + z_d^2 - 2zz_d + \rho^2 \right)^{5/2}} -$$
$$- \frac{15\omega^2 \rho^3 y\left( x^2 + y^2 \right)\pi}{8c^2 \left( R^2 + z_d^2 - 2zz_d + \rho^2 \right)^2} - \frac{\omega^3 \rho x \pi}{c^3}\sqrt{R^2 + z_d^2 - 2zz_d + \rho^2} + \frac{7\omega^3 \rho^3 x\left( x^2 + y^2 \right)\pi}{4c^3 \left( R^2 + z_d^2 - 2zz_d + \rho^2 \right)^{3/2}} .$$

$$\int_0^{2\pi} D\cos\hat{\phi}\,d\phi \approx \frac{\rho x\pi}{R^2+z_d^2-2zz_d+\rho^2}+\frac{3\rho^3 x\left(x^2+y^2\right)\pi}{4\left(R^2+z_d^2-2zz_d+\rho^2\right)^3}-\frac{3\omega\rho^3 y\left(x^2+y^2\right)\pi}{4c\left(R^2+z_d^2-2zz_d+\rho^2\right)^{5/2}}-$$

$$-\frac{15\omega^2\rho^3 x\left(x^2+y^2\right)\pi}{8c^2\left(R^2+z_d^2-2zz_d+\rho^2\right)^2}+\frac{\omega^3\rho y\pi}{c^3}\sqrt{R^2+z_d^2-2zz_d+\rho^2}-\frac{7\omega^3\rho^3 y\left(x^2+y^2\right)\pi}{4c^3\left(R^2+z_d^2-2zz_d+\rho^2\right)^{3/2}}.$$

(37)

From (35) and (37) it follows:

$$A_{dx} \approx -\frac{\mu_0\omega s\rho_{0q}x}{4}\int_0^{\rho_d}\left[\frac{3\omega\rho^2\left(x^2+y^2\right)}{4c\left(R^2+z_d^2-2zz_d+\rho^2\right)^3}-\frac{\omega^3}{c^3}+\frac{7\omega^3\rho^2\left(x^2+y^2\right)}{4c^3\left(R^2+z_d^2-2zz_d+\rho^2\right)^2}\right]\gamma'\rho^3 d\rho-$$

$$-\frac{\mu_0\omega s\rho_{0q}y}{4}\int_0^{\rho_d}\left[1+\frac{3\rho^2\left(x^2+y^2\right)}{4\left(R^2+z_d^2-2zz_d+\rho^2\right)^2}-\frac{15\omega^2\rho^2\left(x^2+y^2\right)}{8c^2\left(R^2+z_d^2-2zz_d+\rho^2\right)}\right]\frac{\gamma'\rho^3 d\rho}{\left(R^2+z_d^2-2zz_d+\rho^2\right)^{3/2}}.$$

$$A_{dy} \approx -\frac{\mu_0\omega s\rho_{0q}y}{4}\int_0^{\rho_d}\left[\frac{3\omega\rho^2\left(x^2+y^2\right)}{4c\left(R^2+z_d^2-2zz_d+\rho^2\right)^3}-\frac{\omega^3}{c^3}+\frac{7\omega^3\rho^2\left(x^2+y^2\right)}{4c^3\left(R^2+z_d^2-2zz_d+\rho^2\right)^2}\right]\gamma'\rho^3 d\rho+$$

$$+\frac{\mu_0\omega s\rho_{0q}x}{4}\int_0^{\rho_d}\left[1+\frac{3\rho^2\left(x^2+y^2\right)}{4\left(R^2+z_d^2-2zz_d+\rho^2\right)^2}-\frac{15\omega^2\rho^2\left(x^2+y^2\right)}{8c^2\left(R^2+z_d^2-2zz_d+\rho^2\right)}\right]\frac{\gamma'\rho^3 d\rho}{\left(R^2+z_d^2-2zz_d+\rho^2\right)^{3/2}}.$$

(38)

Let us substitute the Lorentz factor $\gamma'$ from (27) into (38). Next we will consider the following integrals:

$$I_5 \approx \gamma_c\int_0^{\rho_d}\left[1-\frac{2\pi\eta\rho_0\left(\rho^2+z_d^2\right)}{3c^2}\right]\left[\begin{array}{l}\frac{3\omega\rho^2\left(x^2+y^2\right)}{4c\left(R^2+z_d^2-2zz_d+\rho^2\right)^3}-\frac{\omega^3}{c^3}+\\ +\frac{7\omega^3\rho^2\left(x^2+y^2\right)}{4c^3\left(R^2+z_d^2-2zz_d+\rho^2\right)^2}\end{array}\right]\rho^3 d\rho\,.$$

$$I_6 \approx \gamma_c \int_0^{\rho_d} \left[ \begin{array}{l} 1 - \dfrac{2\pi\eta\rho_0\left(\rho^2 + z_d^2\right)}{3c^2} + \\ + \dfrac{3\rho^2\left(x^2+y^2\right)}{4\left(R^2 + z_d^2 - 2zz_d + \rho^2\right)^2} - \dfrac{15\omega^2\rho^2\left(x^2+y^2\right)}{8c^2\left(R^2 + z_d^2 - 2zz_d + \rho^2\right)} \end{array} \right] \frac{\rho^3\, d\rho}{\left(R^2 + z_d^2 - 2zz_d + \rho^2\right)^{3/2}}. \qquad (39)$$

With the help of (39) expressions (38) are written as follows:

$$A_{dx} \approx -\frac{\mu_0\omega s\rho_{0q}x}{4} I_5 - \frac{\mu_0\omega s\rho_{0q}y}{4} I_6, \qquad A_{dy} \approx -\frac{\mu_0\omega s\rho_{0q}y}{4} I_5 + \frac{\mu_0\omega s\rho_{0q}x}{4} I_6. \qquad (40)$$

After integrating the integrals (39) over the variable $\rho$, in view of the relation $\rho_d = \sqrt{a^2 - z_d^2}$ we obtain:

$$I_5 \approx \gamma_c \left( 1 - \frac{2\pi\eta\rho_0 z_d^2}{3c^2} \right) D_1 - \frac{2\pi\eta\rho_0\gamma_c}{3c^2} D_2,$$

where

$$D_1 = \int_0^{\rho_d} \left[ \frac{3\omega\rho^2\left(x^2+y^2\right)}{4c\left(R^2 + z_d^2 - 2zz_d + \rho^2\right)^3} - \frac{\omega^3}{c^3} + \frac{7\omega^3\rho^2\left(x^2+y^2\right)}{4c^3\left(R^2 + z_d^2 - 2zz_d + \rho^2\right)^2} \right] \rho^3 d\rho.$$

$$D_2 = \int_0^{\rho_d} \left[ \frac{3\omega\rho^2\left(x^2+y^2\right)}{4c\left(R^2 + z_d^2 - 2zz_d + \rho^2\right)^3} - \frac{\omega^3}{c^3} + \frac{7\omega^3\rho^2\left(x^2+y^2\right)}{4c^3\left(R^2 + z_d^2 - 2zz_d + \rho^2\right)^2} \right] \rho^5 d\rho.$$

Besides

$$D_1 \approx \frac{3\omega\left(x^2+y^2\right)}{4c\left(R^2 + z_d^2 - 2zz_d\right)^3} \left[ \frac{\left(a^2 - z_d^2\right)^3}{6} - \frac{3\left(a^2 - z_d^2\right)^4}{8\left(R^2 + z_d^2 - 2zz_d\right)} + \frac{3\left(a^2 - z_d^2\right)^5}{5\left(R^2 + z_d^2 - 2zz_d\right)^2} \right] -$$
$$- \frac{\omega^3}{c^3} \frac{\left(a^2 - z_d^2\right)^2}{4} + \frac{7\omega^3\left(x^2+y^2\right)}{4c^3\left(R^2 + z_d^2 - 2zz_d\right)^2} \left[ \frac{\left(a^2 - z_d^2\right)^3}{6} - \frac{\left(a^2 - z_d^2\right)^4}{4\left(R^2 + z_d^2 - 2zz_d\right)} + \frac{3\left(a^2 - z_d^2\right)^5}{10\left(R^2 + z_d^2 - 2zz_d\right)^2} \right].$$

$$D_2 \approx \frac{3\omega\left(x^2+y^2\right)}{4c\left(R^2+z_d^2-2zz_d\right)^3}\left[\frac{\left(a^2-z_d^2\right)^4}{8}-\frac{3\left(a^2-z_d^2\right)^5}{10\left(R^2+z_d^2-2zz_d\right)}+\frac{\left(a^2-z_d^2\right)^6}{2\left(R^2+z_d^2-2zz_d\right)^2}\right]-$$

$$-\frac{\omega^3}{c^3}\frac{\left(a^2-z_d^2\right)^3}{4}+\frac{7\omega^3\left(x^2+y^2\right)}{4c^3\left(R^2+z_d^2-2zz_d\right)^2}\left[\frac{\left(a^2-z_d^2\right)^4}{8}-\frac{\left(a^2-z_d^2\right)^5}{5\left(R^2+z_d^2-2zz_d\right)}+\frac{\left(a^2-z_d^2\right)^6}{4\left(R^2+z_d^2-2zz_d\right)^2}\right].$$

(41)

In addition, we have

$$I_6 = D_3 + D_4 + D_5 + D_6,$$

where

$$D_3 = \left(\gamma_c - \frac{2\pi\eta\rho_0 z_d^2\gamma_c}{3c^2}\right)\int_0^{\rho_d}\frac{\rho^3\,d\rho}{\left(R^2+z_d^2-2zz_d+\rho^2\right)^{3/2}} =$$

$$=\left(\gamma_c - \frac{2\pi\eta\rho_0 z_d^2\gamma_c}{3c^2}\right)\left(\sqrt{R^2+a^2-2zz_d}+\frac{R^2+z_d^2-2zz_d}{\sqrt{R^2+a^2-2zz_d}}-2\sqrt{R^2+z_d^2-2zz_d}\right).$$

$$D_4 = -\frac{2\pi\eta\rho_0\gamma_c}{3c^2}\int_0^{\rho_d}\frac{\rho^5\,d\rho}{\left(R^2+z_d^2-2zz_d+\rho^2\right)^{3/2}} =$$

$$=-\frac{2\pi\eta\rho_0\gamma_c}{3c^2}\left[\begin{array}{l}\frac{\left(R^2+a^2-2zz_d\right)^{3/2}}{3}+\frac{8\left(R^2+z_d^2-2zz_d\right)^{3/2}}{3}- \\ -2\left(R^2+z_d^2-2zz_d\right)\sqrt{R^2+a^2-2zz_d}-\frac{\left(R^2+z_d^2-2zz_d\right)^2}{\sqrt{R^2+a^2-2zz_d}}\end{array}\right].$$

$$D_5 = \frac{3\left(x^2+y^2\right)\gamma_c}{4}\int_0^{\rho_d}\frac{\rho^5\,d\rho}{\left(R^2+z_d^2-2zz_d+\rho^2\right)^{7/2}} =$$

$$=\frac{3\left(x^2+y^2\right)\gamma_c}{4}\left[\begin{array}{l}-\frac{1}{\sqrt{R^2+a^2-2zz_d}}+\frac{8}{15\sqrt{R^2+z_d^2-2zz_d}}+\frac{2\left(R^2+z_d^2-2zz_d\right)}{3\left(R^2+a^2-2zz_d\right)^{3/2}}- \\ -\frac{\left(R^2+z_d^2-2zz_d\right)^2}{5\left(R^2+a^2-2zz_d\right)^{5/2}}\end{array}\right].$$

$$D_6 = -\frac{15\omega^2\left(x^2+y^2\right)\gamma_c}{8c^2}\int_0^{\rho_d}\frac{\rho^5\, d\rho}{\left(R^2+z_d^2-2zz_d+\rho^2\right)^{5/2}} =$$

$$= -\frac{15\omega^2\left(x^2+y^2\right)\gamma_c}{8c^2}\left[\begin{array}{l}\sqrt{R^2+a^2-2zz_d}+\dfrac{2\left(R^2+z_d^2-2zz_d\right)}{\sqrt{R^2+a^2-2zz_d}}- \\ -\dfrac{8\sqrt{R^2+z_d^2-2zz_d}}{3}-\dfrac{\left(R^2+z_d^2-2zz_d\right)^2}{3\left(R^2+a^2-2zz_d\right)^{3/2}}\end{array}\right].$$

(42)

Substituting in (40) $s$ with the differential $dz_d$ and integrating over all the disks inside the sphere between $-a$ and $a$, we arrive at the components $A_x$ and $A_y$ of the vector potential from the entire sphere:

$$A_x \approx -\frac{\mu_0\,\omega\,\rho_{0q}\,x}{4}\int_{-a}^{a} I_5\, dz_d - \frac{\mu_0\,\omega\,\rho_{0q}\,y}{4}\int_{-a}^{a} I_6\, dz_d\ .$$

$$A_y \approx -\frac{\mu_0\,\omega\,\rho_{0q}\,y}{4}\int_{-a}^{a} I_5\, dz_d + \frac{\mu_0\,\omega\,\rho_{0q}\,x}{4}\int_{-a}^{a} I_6\, dz_d\ .$$

Here we will take into account that the integrals $I_5$ and $I_6$ in (39) are calculated using the quantities $D_1$, $D_2$, $D_3$, $D_4$, $D_5$ and $D_6$ from (41-42):

$$A_x \approx -\frac{\mu_0\,\omega\,\rho_{0q}\,x\gamma_c}{4}\int_{-a}^{a}\left(1-\frac{2\pi\eta\,\rho_0\, z_d^2}{3c^2}\right)D_1\, dz_d + \frac{\pi\mu_0\eta\,\rho_0\,\omega\,\rho_{0q}\,x\gamma_c}{6c^2}\int_{-a}^{a} D_2\, dz_d -$$
$$-\frac{\mu_0\,\omega\,\rho_{0q}\,y}{4}\int_{-a}^{a}\left(D_3+D_4+D_5+D_6\right)dz_d.$$

$$A_y \approx -\frac{\mu_0\,\omega\,\rho_{0q}\,y\gamma_c}{4}\int_{-a}^{a}\left(1-\frac{2\pi\eta\,\rho_0\, z_d^2}{3c^2}\right)D_1\, dz_d + \frac{\pi\mu_0\eta\,\rho_0\,\omega\,\rho_{0q}\,y\gamma_c}{6c^2}\int_{-a}^{a} D_2\, dz_d +$$
$$+\frac{\mu_0\,\omega\,\rho_{0q}\,x}{4}\int_{-a}^{a}\left(D_3+D_4+D_5+D_6\right)dz_d.$$

(43)

The integrals of the quantities $\left(1-\frac{2\pi\eta\rho_0 z_d^2}{3c^2}\right)D_1$, $D_2$, $D_3$, $D_4$, $D_5$ and $D_6$ over the variable $z_d$ are weakly dependent on $z$ and in the first approximation are equal to:

$$\int_{-a}^{a}\left(1-\frac{2\pi\eta\rho_0 z_d^2}{3c^2}\right)D_1\,dz_d \approx$$

$$\approx \frac{4\omega a^7\left(x^2+y^2\right)}{35cR^6}-\frac{4\omega^3 a^5}{15c^3}+\frac{4\omega^3 a^7\left(x^2+y^2\right)}{15c^3R^4}-\frac{8\pi\eta\omega\rho_0 a^9\left(x^2+y^2\right)}{945c^3R^6}.$$

$$\int_{-a}^{a} D_2\,dz_d \approx \frac{8\omega a^9\left(x^2+y^2\right)}{105cR^6}-\frac{8\omega^3 a^7}{35c^3}+\frac{8\omega^3 a^9\left(x^2+y^2\right)}{45c^3R^4}.$$

$$\int_{-a}^{a} D_3\,dz_d \approx \frac{4a^5\gamma_c}{15R^3}-\frac{8\pi\eta\rho_0 a^7\gamma_c}{315c^2R^3}, \qquad \int_{-a}^{a} D_4\,dz_d \approx -\frac{32\pi\eta\rho_0 a^7\gamma_c}{315c^2R^3}.$$

$$\int_{-a}^{a} D_5\,dz_d \approx \frac{4a^7\left(x^2+y^2\right)\gamma_c}{35R^7}, \qquad \int_{-a}^{a} D_6\,dz_d \approx -\frac{2\omega^2 a^7\left(x^2+y^2\right)\gamma_c}{7c^2R^5}.$$

Substituting these integrals into (43), we find:

$$A_x \approx -\frac{\mu_0\omega^2\rho_{0q}a^3 x\gamma_c}{5c}\left[\frac{a^4\left(x^2+y^2\right)}{7R^6}-\frac{\omega^2a^2}{3c^2}+\frac{\omega^2a^4\left(x^2+y^2\right)}{3c^2R^4}-\frac{2\pi\eta\rho_0 a^6\left(x^2+y^2\right)}{27c^2R^6}\right]-$$

$$-\frac{\mu_0\omega\rho_{0q}a^5 y\gamma_c}{15R^3}\left[1-\frac{10\pi\eta\rho_0 a^2}{21c^2}+\frac{3a^2\left(x^2+y^2\right)}{7R^4}-\frac{15\omega^2a^2\left(x^2+y^2\right)}{14c^2R^2}\right].$$

$$A_y \approx -\frac{\mu_0\omega^2\rho_{0q}a^3 y\gamma_c}{5c}\left[\frac{a^4\left(x^2+y^2\right)}{7R^6}-\frac{\omega^2a^2}{3c^2}+\frac{\omega^2a^4\left(x^2+y^2\right)}{3c^2R^4}-\frac{2\pi\eta\rho_0 a^6\left(x^2+y^2\right)}{27c^2R^6}\right]+$$

$$+\frac{\mu_0\omega\rho_{0q}a^5 x\gamma_c}{15R^3}\left[1-\frac{10\pi\eta\rho_0 a^2}{21c^2}+\frac{3a^2\left(x^2+y^2\right)}{7R^4}-\frac{15\omega^2a^2\left(x^2+y^2\right)}{14c^2R^2}\right].$$

(44)

In view of the approximate nature of our calculations, we should define more precisely all the terms in (44) by substituting the components $A_x$ and $A_y$ of the vector potential into the Laplace equation (10), which has the form $\Delta \mathbf{A}=0$. For this equation to hold, we need to perform simplification in (44), eliminating the small terms and assuming $\frac{x^2+y^2}{R^2}\approx 1$. Previously we used a similar approach, in order to pass on from (32) to expression (33) for the potential. This gives the following expression, which is valid at small $z$:

$$A_x \approx \frac{\mu_0 \omega^4 \rho_{0q} a^5 x \gamma_c}{15c^3} - \frac{\mu_0 \omega \rho_{0q} a^5 y \gamma_c}{15R^3}\left(1-\frac{10\pi\eta\rho_0 a^2}{21c^2}-\frac{15\omega^2 a^2}{14c^2}\right).$$

$$A_y \approx \frac{\mu_0 \omega^4 \rho_{0q} a^5 y \gamma_c}{15c^3} + \frac{\mu_0 \omega \rho_{0q} a^5 x \gamma_c}{15R^3}\left(1-\frac{10\pi\eta\rho_0 a^2}{21c^2}-\frac{15\omega^2 a^2}{14c^2}\right). \qquad (45)$$

Since in (35) $A_{dz}=0$ for each rotating disk inside the sphere, then the component $A_z$ of the vector potential from the entire rotating sphere with charged particles is also equal to zero.

### 2.4. Electric and magnetic fields in the middle zone

The electric field strength $\mathbf{E}$ and the magnetic field induction $\mathbf{B}$ are given by standard formulas:

$$\mathbf{E}=-\nabla\varphi-\frac{\partial \mathbf{A}}{\partial t}, \qquad \mathbf{B}=\nabla\times\mathbf{A}. \qquad (46)$$

Since the sphere rotates at the constant angular velocity $\omega$, the vector potential components in (45) do not depend on time, and then the field $\mathbf{E}$ is defined only by the gradient of the scalar potential $\varphi$. Let us substitute (33) and (45) into (46) and find the fields $\mathbf{E}$ and $\mathbf{B}$, taking into account that $A_z=0$:

$$\mathbf{E}\approx \frac{q_\omega \mathbf{R}}{4\pi\varepsilon_0 R^3}\left(1+\frac{\omega^2 a^2}{10c^2}\right).$$

$$B_x = \frac{\partial A_z}{\partial y} - \frac{\partial A_y}{\partial z} \approx \frac{\mu_0 \omega \rho_{0q} a^5 xz\gamma_c}{5R^5}\left(1 - \frac{10\pi\eta\rho_0 a^2}{21c^2} - \frac{15\omega^2 a^2}{14c^2}\right).$$

$$B_y = \frac{\partial A_x}{\partial z} - \frac{\partial A_z}{\partial x} \approx \frac{\mu_0 \omega \rho_{0q} a^5 yz\gamma_c}{5R^5}\left(1 - \frac{10\pi\eta\rho_0 a^2}{21c^2} - \frac{15\omega^2 a^2}{14c^2}\right).$$

$$B_z = \frac{\partial A_y}{\partial x} - \frac{\partial A_x}{\partial y} \approx \frac{\mu_0 \omega \rho_{0q} a^5 \gamma_c \left(2R^2 - 3x^2 - 3y^2\right)}{15R^5}\left(1 - \frac{10\pi\eta\rho_0 a^2}{21c^2} - \frac{15\omega^2 a^2}{14c^2}\right).$$

(47)

Since we simplified (44) and used for the vector potential (45), (47) contains only the dipole component of the magnetic field.

In the special theory of relativity, the wave equations are valid for the electric and magnetic fields [15]:

$$\partial_\beta \partial^\beta \mathbf{E} = \frac{1}{c^2}\frac{\partial^2 \mathbf{E}}{\partial t^2} - \Delta \mathbf{E} = -\frac{1}{\varepsilon_0}\nabla(\gamma\rho_{0q}) - \mu_0 \frac{\partial \mathbf{j}}{\partial t}, \qquad \partial_\beta \partial^\beta \mathbf{B} = \frac{1}{c^2}\frac{\partial^2 \mathbf{B}}{\partial t^2} - \Delta \mathbf{B} = \mu_0 \nabla \times \mathbf{j}.$$

Since there are no charges or currents outside the rotating charged sphere, the right-hand side of the wave equations becomes equal to zero. In addition, at the constant velocity of rotation, $\mathbf{E}$ and $\mathbf{B}$ do not depend on time. As a result, the wave equations for the fields turn into Laplace equations:

$$\Delta \mathbf{E} = 0, \qquad \Delta \mathbf{B} = 0. \tag{48}$$

By directly substituting the components of the electric field $\mathbf{E}$ and the magnetic field $\mathbf{B}$ from (47) into (48), we can make sure that the fields in the middle zone satisfy the Laplace equation.

### 2.5. Scalar potential in the far zone

As the condition for the far zone we can consider the conditions $R >> a$, $\frac{\omega \hat{R}_P}{c} \approx 1$. Since $\hat{\phi} = \phi - \frac{\omega \hat{R}_P}{c}$, in this case we can write:

$$\cos\hat{\phi} = \cos\left(\phi - \phi_P\right), \qquad \sin\hat{\phi} = \sin\left(\phi - \phi_P\right), \tag{49}$$

where, in view of (13), the angle $\phi_P = \frac{\omega \hat{R}_P}{c} = \frac{\omega}{c}\sqrt{R^2 + z_d^2 - 2zz_d + \rho^2 - 2\rho x\cos\hat{\phi} - 2\rho y\sin\hat{\phi}} \approx \frac{\omega R}{c}$.

Substitution of (49) into (16) gives the following:

$$\varphi_d = \frac{s\rho_{0q}}{4\pi\varepsilon_0}\int_0^{2\pi}\int_0^{\rho_d}\frac{\gamma'\rho\,d\rho\,d\phi}{\hat{R}_P + \frac{\omega\rho x\sin\left(\phi-\phi_P\right)}{c} - \frac{\omega\rho y\cos\left(\phi-\phi_P\right)}{c}}. \tag{50}$$

where $\hat{R}_P = \sqrt{R^2 + z_d^2 - 2zz_d + \rho^2 - 2\rho x\cos\left(\phi-\phi_P\right) - 2\rho y\sin\left(\phi-\phi_P\right)}$.

Let us take into account the following transformations for the expression under the integral sign:

$$H = \frac{1}{\hat{R}_P + \frac{\omega\rho x\sin\left(\phi-\phi_P\right)}{c} - \frac{\omega\rho y\cos\left(\phi-\phi_P\right)}{c}} =$$

$$= \frac{\left(R^2 + z_d^2 - 2zz_d + \rho^2\right)^{-1/2}}{\sqrt{1 - \frac{2\rho\left[x\cos\left(\phi-\phi_P\right) + y\sin\left(\phi-\phi_P\right)\right]}{R^2 + z_d^2 - 2zz_d + \rho^2}} + \frac{\omega\rho\left[x\sin\left(\phi-\phi_P\right) - y\cos\left(\phi-\phi_P\right)\right]}{c\sqrt{R^2 + z_d^2 - 2zz_d + \rho^2}}}.$$

In this expression, we use the rule for expanding the square root in the form $\sqrt{1-\delta} \approx 1 - \frac{\delta}{2} - \frac{\delta^2}{8}$ and an approximate expression $\frac{1}{1 - \frac{\delta}{2} - \frac{\delta^2}{8} + \gamma} \approx 1 + \frac{\delta}{2} + \frac{3\delta^2}{8} - \gamma - \gamma\delta\left(1+\delta\right) + \gamma^2$:

$$H \approx \frac{\left( \begin{array}{l} 1+\frac{\rho\left[x\cos(\phi-\phi_P)+y\sin(\phi-\phi_P)\right]}{R^2+z_d^2-2zz_d+\rho^2}+\frac{3\rho^2\left[x\cos(\phi-\phi_P)+y\sin(\phi-\phi_P)\right]^2}{2\left(R^2+z_d^2-2zz_d+\rho^2\right)^2}- \\ -\frac{\omega\rho\left[x\sin(\phi-\phi_P)-y\cos(\phi-\phi_P)\right]}{c\sqrt{R^2+z_d^2-2zz_d+\rho^2}}+\frac{\omega^2\rho^2\left[x\sin(\phi-\phi_P)-y\cos(\phi-\phi_P)\right]^2}{c^2\left(R^2+z_d^2-2zz_d+\rho^2\right)}- \\ -\frac{2\omega\rho^2\left[x\sin(\phi-\phi_P)-y\cos(\phi-\phi_P)\right]\left[x\cos(\phi-\phi_P)+y\sin(\phi-\phi_P)\right]}{c\left(R^2+z_d^2-2zz_d+\rho^2\right)^{3/2}} \end{array} \right)}{\sqrt{R^2+z_d^2-2zz_d+\rho^2}}. \tag{51}$$

In view of (51), for the potential (50) we can write the following:

$$\varphi_d \approx \frac{s\rho_{0q}}{4\pi\varepsilon_0}\int_0^{2\pi}\int_0^{\rho_d} H\gamma'\rho\, d\rho\, d\phi. \tag{52}$$

As the distance $R$ increases, the angle $\phi_P = \frac{\omega \hat{R}_P}{c} \approx \frac{\omega R}{c}$ can first reach the value $\frac{\pi}{2}$, then $\pi$, $\frac{3\pi}{2}$, $2\pi$, etc. In the general case, the angle $\phi_P$ will pass through the values $\frac{k\pi}{2}$, where $k = 1,2,3...$

Let us integrate the quantity $H$ in (52) over the angle $\phi$, assuming the angle $\phi_P$ to be constant and almost independent of $\phi$. Taking into account that the integrals of $\cos(\phi-\phi_P)$, $\sin(\phi-\phi_P)$ and $\sin(\phi-\phi_P)\cos(\phi-\phi_P)$ between the limits of 0 and $2\pi$ are equal to zero, we find:

$$\varphi_d \approx \frac{s\rho_{0q}}{2\varepsilon_0}\int_0^{\rho_d} \frac{\left[1+\frac{\omega^2\rho^2\left(x^2+y^2\right)}{2c^2\left(R^2+z_d^2-2zz_d+\rho^2\right)}+\frac{3\rho^2\left(x^2+y^2\right)}{4\left(R^2+z_d^2-2zz_d+\rho^2\right)^2}\right]\gamma'\rho\, d\rho}{\sqrt{R^2+z_d^2-2zz_d+\rho^2}}. \tag{53}$$

If we substitute the expression for $\gamma'$ from (27) into (53), then we will see that the potential can be represented in the form $\varphi_d \approx \frac{s\rho_{0q}\gamma_c}{2\varepsilon_0}\left(I_1+I_2+2I_3+I_4\right)$, where the integrals $I_1$, $I_2$, $I_3$ and $I_4$ were found in (28).

The sum of the potentials $\varphi_d$ of all the sphere's layers gives the sought-for sphere potential. Assuming $s = dz_d$ and substituting the sum of the layers' potentials with the integral over the variable $z_d$, for the sphere potential we can write:

$$\begin{aligned}\varphi &\approx \frac{\rho_{0q}\gamma_c}{2\varepsilon_0}\int_{-a}^{a}\left(I_1+I_2+2I_3+2I_4\right)dz_d \approx \\ &\approx \frac{\rho_{0q}a^3\gamma_c}{3\varepsilon_0 R}\left[1-\frac{2\pi\eta\rho_0 a^2}{5c^2}+\frac{3a^2\left(x^2+y^2\right)}{10R^4}+\frac{\omega^2a^2\left(x^2+y^2\right)}{5c^2R^2}\right].\end{aligned} \tag{54}$$

The scalar potential (54) in the far zone differs from the potential (29) in the middle zone in the fact that in (54) the last term in the square brackets is twice as large.

In (54), we can substitute (31) and express the potential in terms of the charge $q_\omega$. To ensure that the potential corresponds to the Laplace equation, in (54) we will eliminate the small term $\frac{3a^2\left(x^2+y^2\right)}{10R^4}$ and assume that $x^2+y^2 \approx R^2$, which is true at small $z$. As a result we obtain the following:

$$\varphi \approx \frac{q_\omega}{4\pi\varepsilon_0 R}\left(1+\frac{\omega^2a^2}{5c^2}\right). \tag{55}$$

We suppose that the small difference between the potentials in (55) and in (33) is associated with the fact that the solutions for these potentials were obtained in two different ways and with different degrees of approximation.

### 2.6. Vector potential in the far zone

We will transform (18) using (51) in the same way as potential (50) was transformed, and we will also take into account (49). Then for the vector potential components of the rotating disk we find the following:

$$A_{dx} \approx -\frac{\mu_0 \omega s \rho_{0q}}{4\pi} \int_0^{2\pi} \int_0^{\rho_d} H \gamma' \sin(\phi - \phi_P) \rho^2 d\rho d\phi .$$

$$A_{dy} \approx \frac{\mu_0 \omega s \rho_{0q}}{4\pi} \int_0^{2\pi} \int_0^{\rho_d} H \gamma' \cos(\phi - \phi_P) \rho^2 d\rho d\phi .$$

At large distances, we may neglect the change in the angle $\phi_P = \frac{\omega \hat{R}_P}{c} \approx \frac{\omega R}{c}$ when integrating over the angle $\phi$ and consider $\phi_P$ a constant. This makes it easier to integrate components $A_{dx}$ and $A_{dy}$. Taking into account the expression for $H$ from (51), we find:

$$A_{dx} \approx \frac{\mu_0 \omega^2 s \rho_{0q} x}{4c} \int_0^{\rho_d} \frac{\gamma' \rho^3 d\rho}{R^2 + z_d^2 - 2z z_d + \rho^2} - \frac{\mu_0 \omega s \rho_{0q} y}{4} \int_0^{\rho_d} \frac{\gamma' \rho^3 d\rho}{\left(R^2 + z_d^2 - 2z z_d + \rho^2\right)^{3/2}} .$$

$$A_{dy} \approx \frac{\mu_0 \omega^2 s \rho_{0q} y}{4c} \int_0^{\rho_d} \frac{\gamma' \rho^3 d\rho}{R^2 + z_d^2 - 2z z_d + \rho^2} + \frac{\mu_0 \omega s \rho_{0q} x}{4} \int_0^{\rho_d} \frac{\gamma' \rho^3 d\rho}{\left(R^2 + z_d^2 - 2z z_d + \rho^2\right)^{3/2}} .$$

(56)

If we substitute the Lorentz factor $\gamma'$ from (27) into (56), the following integrals appear:

$$I_7 = \gamma_c \int_0^{\rho_d} \left[ 1 - \frac{2\pi \eta \rho_0 \left(\rho^2 + z_d^2\right)}{3c^2} \right] \frac{\rho^3 d\rho}{R^2 + z_d^2 - 2z z_d + \rho^2} .$$

$$I_8 = \gamma_c \int_0^{\rho_d} \left[ 1 - \frac{2\pi \eta \rho_0 \left(\rho^2 + z_d^2\right)}{3c^2} \right] \frac{\rho^3 d\rho}{\left(R^2 + z_d^2 - 2z z_d + \rho^2\right)^{3/2}} .$$

Using the integrals $I_7$ and $I_8$ we can write (56) as follows:

$$A_{dx} \approx \frac{\mu_0 \omega^2 s \rho_{0q} x}{4c} I_7 - \frac{\mu_0 \omega s \rho_{0q} y}{4} I_8 , \qquad A_{dy} \approx \frac{\mu_0 \omega^2 s \rho_{0q} y}{4c} I_7 + \frac{\mu_0 \omega s \rho_{0q} x}{4} I_8 . \qquad (57)$$

Let us calculate the integrals $I_7$ and $I_8$ taking into account the relation $\rho_d = \sqrt{a^2 - z_d^2}$, expanding the denominators into series by the rule $\frac{1}{1+\delta} \approx 1 - \delta + \delta^2$, where $\delta = \frac{\rho^2}{R^2 + z_d^2 - 2zz_d}$

:

$$I_7 \approx \frac{\left(a^2 - z_d^2\right)^2}{4\left(R^2 + z_d^2 - 2zz_d\right)}\left(\gamma_c - \frac{2\pi\eta\rho_0 z_d^2\gamma_c}{3c^2}\right)\left[1 - \frac{2\left(a^2 - z_d^2\right)}{3\left(R^2 + z_d^2 - 2zz_d\right)} + \frac{\left(a^2 - z_d^2\right)^2}{2\left(R^2 + z_d^2 - 2zz_d\right)^2}\right] -$$
$$- \frac{\pi\eta\rho_0\gamma_c\left(a^2 - z_d^2\right)^3}{9c^2\left(R^2 + z_d^2 - 2zz_d\right)}\left[1 - \frac{3\left(a^2 - z_d^2\right)}{4\left(R^2 + z_d^2 - 2zz_d\right)} + \frac{3\left(a^2 - z_d^2\right)^2}{5\left(R^2 + z_d^2 - 2zz_d\right)^2}\right].$$

$$I_8 = \left(\gamma_c - \frac{2\pi\eta\rho_0 z_d^2\gamma_c}{3c^2}\right)\left[\sqrt{R^2 + a^2 - 2zz_d} + \frac{R^2 + z_d^2 - 2zz_d}{\sqrt{R^2 + a^2 - 2zz_d}} - 2\sqrt{R^2 + z_d^2 - 2zz_d}\right] -$$
$$- \frac{2\pi\eta\rho_0\gamma_c}{3c^2}\left[\begin{array}{l} \frac{\left(R^2 + a^2 - 2zz_d\right)^{3/2}}{3} - 2\left(R^2 + z_d^2 - 2zz_d\right)\sqrt{R^2 + a^2 - 2zz_d} - \\ - \frac{\left(R^2 + z_d^2 - 2zz_d\right)^2}{\sqrt{R^2 + a^2 - 2zz_d}} + \frac{8\left(R^2 + z_d^2 - 2zz_d\right)^{3/2}}{3} \end{array}\right].$$

(58)

The quantities $A_{dx}$ and $A_{dy}$ are the components of the vector potential from one thin disk. To pass on to the corresponding components of the potential from the entire sphere, in (57) it is necessary to set $s = dz_d$ and integrate over the variable $z_d$ that specifies the position of the disks inside the sphere on the axis $OZ$:

$$A_x \approx \frac{\mu_0\omega^2\rho_{0q}x}{4c}\int_{-a}^{a} I_7\, dz_d - \frac{\mu_0\omega\rho_{0q}y}{4}\int_{-a}^{a} I_8\, dz_d\,,$$

$$A_y \approx \frac{\mu_0\omega^2\rho_{0q}y}{4c}\int_{-a}^{a} I_7\, dz_d + \frac{\mu_0\omega\rho_{0q}x}{4}\int_{-a}^{a} I_8\, dz_d\,. \qquad (59)$$

Substitution of (58) into (59) and subsequent integration over the variable $z_d$ give the following:

$$A_x \approx \frac{\mu_0 \omega^2 \rho_{0q} a^5 x\gamma_c}{15cR^2}\left(1-\frac{10\pi\eta\,\rho_0 a^2}{21c^2}\right)-\frac{\mu_0 \omega\,\rho_{0q} a^5 y\gamma_c}{15R^3}\left(1-\frac{10\pi\eta\,\rho_0 a^2}{21c^2}\right).$$

$$A_y \approx \frac{\mu_0 \omega^2 \rho_{0q} a^5 y\gamma_c}{15cR^2}\left(1-\frac{10\pi\eta\,\rho_0 a^2}{21c^2}\right)+\frac{\mu_0 \omega\,\rho_{0q} a^5 x\gamma_c}{15R^3}\left(1-\frac{10\pi\eta\,\rho_0 a^2}{21c^2}\right).$$

(60)

Here, only the last terms containing $R^3$ in the denominator exactly satisfy the Laplace equation. As for the first terms, the far zone condition $\frac{\omega R}{c} \approx 1$ can be taken into account in them. This gives the following expressions

$$A_x \approx \frac{\mu_0 \omega\rho_{0q} a^5 x\gamma_c}{15R^3}\left(1-\frac{10\pi\eta\,\rho_0 a^2}{21c^2}\right)-\frac{\mu_0 \omega\,\rho_{0q} a^5 y\gamma_c}{15R^3}\left(1-\frac{10\pi\eta\,\rho_0 a^2}{21c^2}\right),$$

$$A_y \approx \frac{\mu_0 \omega\rho_{0q} a^5 y\gamma_c}{15R^3}\left(1-\frac{10\pi\eta\,\rho_0 a^2}{21c^2}\right)+\frac{\mu_0 \omega\,\rho_{0q} a^5 x\gamma_c}{15R^3}\left(1-\frac{10\pi\eta\,\rho_0 a^2}{21c^2}\right),$$

that satisfy the Laplace equation. The component $A_z = 0$, and therefore $A_z$ automatically satisfies the Laplace equation.

### 2.7. The electric and magnetic fields in the far zone

In order to find the fields $\mathbf{E}$ and $\mathbf{B}$ it is necessary to substitute (55) and (60) into (46):

$$\mathbf{E} \approx \frac{q_\omega \mathbf{R}}{4\pi\varepsilon_0 R^3}\left(1+\frac{\omega^2 a^2}{5c^2}\right).$$

$$B_x = \frac{\partial A_z}{\partial y}-\frac{\partial A_y}{\partial z} \approx \frac{2\mu_0 \omega^2 \rho_{0q} a^5 yz\gamma_c}{15cR^4}\left(1-\frac{10\pi\eta\,\rho_0 a^2}{21c^2}\right)+\frac{\mu_0 \omega\,\rho_{0q} a^5 xz\gamma_c}{5R^5}\left(1-\frac{10\pi\eta\,\rho_0 a^2}{21c^2}\right).$$

$$B_y = \frac{\partial A_x}{\partial z} - \frac{\partial A_z}{\partial x} \approx -\frac{2\mu_0 \omega^2 \rho_{0q} a^5 xz\gamma_c}{15cR^4}\left(1 - \frac{10\pi\eta\rho_0 a^2}{21c^2}\right) + \frac{\mu_0 \omega \rho_{0q} a^5 yz\gamma_c}{5R^5}\left(1 - \frac{10\pi\eta\rho_0 a^2}{21c^2}\right).$$

$$B_z = \frac{\partial A_y}{\partial x} - \frac{\partial A_x}{\partial y} \approx \frac{\mu_0 \omega \rho_{0q} a^5 \gamma_c \left(2R^2 - 3x^2 - 3y^2\right)}{15R^5}\left(1 - \frac{10\pi\eta\rho_0 a^2}{21c^2}\right). \qquad (61)$$

The fields in (61) differ insignificantly from the fields in (47) in the middle zone due to the small additions proportional to the value $\frac{\omega^2 a^2}{c^2}$. This difference can be considered the consequence of the fact that during calculations different methods of obtaining an approximate solution were used. In addition, a rotational component of the magnetic field appears in the first terms in $B_x$ and $B_y$ in (61).

## 2.8. Scalar potential in the near zone

In the near zone the conditions $R \ge a$, $\frac{\omega \hat{R}_P}{c} << 1$ are met, so that the point $P$, where the potential is determined, is not far from the sphere. We can start with expression (21) for the potential $\varphi_d$, generated by a thin disk-shaped layer inside the sphere, located on the axis $OZ$ at the height $z_d$. For the near zone, we can assume that the early time point $\hat{t} = t - \frac{\hat{R}_P}{c}$ is approximately equal to $\hat{t} \approx t - \frac{R_P}{c}$. In this case, the quantity $R_P$ in (11) differs little from $\hat{R}_P$ in (13), since their difference is associated with a small difference between the angle $\phi$ and the angle $\hat{\phi} = \phi - \frac{\omega \hat{R}_P}{c}$. Therefore, the quantity $\hat{R}_P$ in the denominator in (21) can be substituted with $R_P$.

The quantity $R_P$ is the distance from the integration point inside the sphere to the observation point $P$. Further, we will assume that the point $P$ is located outside the sphere and the condition $R_P >> \frac{\omega \rho \left(x \sin\phi - y \cos\phi\right)}{c}$ is met. This allows us to expand the root in (21) so as to distinguish a small term containing the square of the speed of light:

$$\sqrt{R^2+z_d^2-2zz_d+\rho^2-2\rho x\cos\phi-2\rho y\sin\phi+\frac{\omega^2\rho^2\left(x\sin\phi-y\cos\phi\right)^2}{c^2}}\approx$$

$$\approx R_P+\frac{\omega^2\rho^2\left(x\sin\phi-y\cos\phi\right)^2}{2c^2R_P}.$$

Here, the quantity $R_P$ represents the square root and corresponds to (11):

$$R_P=\sqrt{R^2+z_d^2-2zz_d+\rho^2-2\rho x\cos\phi-2\rho y\sin\phi}. \qquad (62)$$

Now the denominator in (21) can be transformed by the rule $\frac{1}{R_P+\delta}\approx\frac{1}{R_P}\left(1-\frac{\delta}{R_P}\right)$. The potential $\varphi_d$ is generated by one layer in the form of a thin disk of radius $\rho_d=\sqrt{a^2-z_d^2}$. The total potential of the sphere is the sum of the potentials of all the layers, and this sum, in view of the equality $s=d\,z_d$, can be substituted with the integral:

$$\varphi\approx\frac{\rho_{0q}}{4\pi\varepsilon_0}\int_{-a}^{a}\int_{0}^{\rho_d}\int_{0}^{2\pi}\left[1-\frac{\omega^2\rho^2\left(x\sin\phi-y\cos\phi\right)^2}{2c^2R_P^2}+\frac{\omega^2\rho x}{c^2}\cos\phi+\frac{\omega^2\rho y}{c^2}\sin\phi\right]\frac{\gamma'\rho\,d\rho\,d\phi\,d\,z_d}{R_P}. \qquad (63)$$

In (63) the expression in square brackets depends on the angle $\phi$, as well as $R_P$ according to (62). When integrating over the angle, we need four integrals:

$$I_9=\int_0^{2\pi}\frac{d\phi}{R_P}=\int_0^{2\pi}\frac{d\phi}{\sqrt{R^2+z_d^2-2zz_d+\rho^2-2\rho x\cos\phi-2\rho y\sin\phi}}.$$

$$I_{10}=\int_0^{2\pi}\frac{\left(x\sin\phi-y\cos\phi\right)^2}{\left(R^2+z_d^2-2zz_d+\rho^2-2\rho x\cos\phi-2\rho y\sin\phi\right)^{3/2}}d\phi.$$

$$I_{11}=\int_0^{2\pi}\frac{\cos\phi}{\sqrt{R^2+z_d^2-2zz_d+\rho^2-2\rho x\cos\phi-2\rho y\sin\phi}}d\phi.$$

$$I_{12} = \int_0^{2\pi} \frac{\sin\phi}{\sqrt{R^2 + z_d^2 - 2zz_d + \rho^2 - 2\rho x\cos\phi - 2\rho y\sin\phi}} d\phi . \tag{64}$$

Using the notation $x = \sqrt{x^2 + y^2}\cos\psi$, $y = \sqrt{x^2 + y^2}\sin\psi$, and then making substitution $\phi - \psi = \pi + 2\gamma$, $\cos(\phi - \psi) = -\cos 2\gamma = 2\sin^2\gamma - 1$, we transform the integral $I_9$:

$$\begin{aligned} I_9 &= \frac{2}{\sqrt{R^2 + z_d^2 - 2zz_d + \rho^2 + 2\rho\sqrt{x^2 + y^2}}} \int_{\frac{-\pi-\psi}{2}}^{\frac{\pi-\psi}{2}} \frac{d\gamma}{\sqrt{1 - k^2\sin^2\gamma}} = \\ &= \frac{4}{\sqrt{R^2 + z_d^2 - 2zz_d + \rho^2 + 2\rho\sqrt{x^2 + y^2}}} F\left(k, \frac{\pi}{2}\right). \end{aligned} \tag{65}$$

Here $k^2 = \dfrac{4\rho\sqrt{x^2 + y^2}}{R^2 + z_d^2 - 2zz_d + \rho^2 + 2\rho\sqrt{x^2 + y^2}}$, $F\left(k, \dfrac{\pi}{2}\right) = \displaystyle\int_0^{\pi/2} \frac{d\gamma}{\sqrt{1 - k^2\sin^2\gamma}}$ is the complete normal elliptic integral of the first kind.

The integral $I_{10}$ can be expressed in a similar way:

$$\begin{aligned} I_{10} = &-\left.\frac{\sqrt{x^2 + y^2}\sin(\phi - \psi)}{\rho\sqrt{R^2 + z_d^2 - 2zz_d + \rho^2 - 2\rho\sqrt{x^2 + y^2}\cos(\phi - \psi)}}\right|_{\phi=0}^{2\pi} - \\ &-\frac{1}{\rho^2} \int_{\frac{-\pi-\psi}{2}}^{\frac{\pi-\psi}{2}} \frac{\left[2\rho\sqrt{x^2 + y^2}\cos 2\gamma + \left(R^2 + z_d^2 - 2zz_d + \rho^2\right) - \left(R^2 + z_d^2 - 2zz_d + \rho^2\right)\right] d\gamma}{\sqrt{R^2 + z_d^2 - 2zz_d + \rho^2 + 2\rho\sqrt{x^2 + y^2}\cos 2\gamma}} . \end{aligned}$$

The first term on the right-hand side vanishes upon substituting the limits of integration $\phi = 0$ and $\phi = 2\pi$. Taking into account the relation $\cos 2\gamma = 1 - 2\sin^2\gamma$, we have:

$$I_{10}=-\frac{\sqrt{R^2+z_d^2-2zz_d+\rho^2+2\rho\sqrt{x^2+y^2}}}{\rho^2}\int_{\frac{-\pi-\psi}{2}}^{\frac{\pi-\psi}{2}}\sqrt{1-k^2\sin^2\gamma}\,d\gamma+$$

$$+\frac{\left(R^2+z_d^2-2zz_d+\rho^2\right)}{\rho^2\sqrt{R^2+z_d^2-2zz_d+\rho^2+2\rho\sqrt{x^2+y^2}}}\int_{\frac{-\pi-\psi}{2}}^{\frac{\pi-\psi}{2}}\frac{d\gamma}{\sqrt{1-k^2\sin^2\gamma}}=$$

$$=-\frac{2\sqrt{R^2+z_d^2-2zz_d+\rho^2+2\rho\sqrt{x^2+y^2}}\,E\left(k,\frac{\pi}{2}\right)}{\rho^2}+\frac{2\left(R^2+z_d^2-2zz_d+\rho^2\right)F\left(k,\frac{\pi}{2}\right)}{\rho^2\sqrt{R^2+z_d^2-2zz_d+\rho^2+2\rho\sqrt{x^2+y^2}}}. \quad (66)$$

Here $E\left(k,\frac{\pi}{2}\right)=\int_0^{\pi/2}\sqrt{1-k^2\sin^2\gamma}\,d\gamma$ is a complete normal elliptic integral of the second kind. Transformation of the integral $I_{11}$ gives the following:

$$I_{11}=-\left.\frac{\sin\psi\sqrt{R^2+z_d^2-2zz_d+\rho^2+2\rho\sqrt{x^2+y^2}\cos\left(\phi-\psi-\pi\right)}}{\rho\sqrt{x^2+y^2}}\right|_{\phi=0}^{2\pi}-$$

$$-\frac{\cos\psi}{\rho\sqrt{x^2+y^2}}\int_{\frac{-\pi-\psi}{2}}^{\frac{\pi-\psi}{2}}\frac{\left[2\rho\sqrt{x^2+y^2}\cos 2\gamma+\left(R^2+z_d^2-2zz_d+\rho^2\right)-\left(R^2+z_d^2-2zz_d+\rho^2\right)\right]d\gamma}{\sqrt{R^2+z_d^2-2zz_d+\rho^2+2\rho\sqrt{x^2+y^2}\cos 2\gamma}}.$$

The first term on the right-hand side is equal to zero, as in case of the integral $I_{10}$. The second term on the right-hand side, in view of the relations $x=\sqrt{x^2+y^2}\cos\psi$, $y=\sqrt{x^2+y^2}\sin\psi$, is transformed so that $I_{11}$ is expressed in terms of elliptic integrals:

$$I_{11}=-\frac{2x\sqrt{R^2+z_d^2-2zz_d+\rho^2+2\rho\sqrt{x^2+y^2}}}{\rho\left(x^2+y^2\right)}E\left(k,\frac{\pi}{2}\right)+$$
$$+\frac{2x\left(R^2+z_d^2-2zz_d+\rho^2\right)}{\rho\left(x^2+y^2\right)\sqrt{R^2+z_d^2-2zz_d+\rho^2+2\rho\sqrt{x^2+y^2}}}F\left(k,\frac{\pi}{2}\right). \quad (67)$$

Similarly to $I_{11}$, for $I_{12}$ we obtain the following:

$$I_{12} = -\frac{2y\sqrt{R^2 + z_d^2 - 2zz_d + \rho^2 + 2\rho\sqrt{x^2+y^2}}}{\rho\left(x^2+y^2\right)} E\left(k, \frac{\pi}{2}\right) + \\ + \frac{2y\left(R^2 + z_d^2 - 2zz_d + \rho^2\right)}{\rho\left(x^2+y^2\right)\sqrt{R^2 + z_d^2 - 2zz_d + \rho^2 + 2\rho\sqrt{x^2+y^2}}} F\left(k, \frac{\pi}{2}\right). \tag{68}$$

Taking into account (64), as well as (27) for $\gamma'$, (63) will be written as follows:

$$\varphi \approx \frac{\rho_{0q}\,\gamma_c}{4\pi\varepsilon_0} \int_{-a}^{a} \int_{0}^{\rho_d} \left\{ \begin{array}{l} \left[1 - \dfrac{2\pi\eta\,\rho_0\left(\rho^2 + z_d^2\right)}{3c^2}\right] I_9 - \dfrac{\omega^2\rho^2}{2c^2} I_{10} + \\ + \dfrac{\omega^2 \rho x}{c^2} I_{11} + \dfrac{\omega^2 \rho y}{c^2} I_{12} \end{array} \right\} \rho\, d\rho\, dz_d \ .$$

This expression shows that we need to calculate the integrals $\int_0^{\rho_d} I_9\, \rho\, d\rho$, $\int_0^{\rho_d} I_9\, \rho^3\, d\rho$, $\int_0^{\rho_d} I_{10}\, \rho^3\, d\rho$, $\int_0^{\rho_d} I_{11}\, \rho^2\, d\rho$, $\int_0^{\rho_d} I_{12}\, \rho^2\, d\rho$. To do this, it is necessary to represent the quantities $I_9$, $I_{10}$, $I_{11}$ and $I_{12}$ so that the variable $\rho$ appears in them in an explicit form. For this purpose, we will expand the elliptic integrals $E\left(k, \frac{\pi}{2}\right)$ and $F\left(k, \frac{\pi}{2}\right)$ in (65-68) into series by the standard formulas:

$$F\left(k, \frac{\pi}{2}\right) = \frac{\pi}{2}\left(1 + \frac{k^2}{4} + \frac{9k^4}{64} ...\right) = \frac{\pi}{2} \sum_{n=0}^{\infty} \left(\frac{(2n)!}{2^{2n} n!^2}\right)^2 k^{2n}.$$

$$E\left(k, \frac{\pi}{2}\right) = \frac{\pi}{2}\left(1 - \frac{k^2}{4} - \frac{3k^4}{64} ...\right) = \frac{\pi}{2} \sum_{n=0}^{\infty} \left(\frac{(2n)!}{2^{2n} n!^2}\right)^2 \frac{k^{2n}}{1-2n}. \tag{69}$$

If in (69) we take into account the first two expansion terms $F\left(k, \frac{\pi}{2}\right)$ and substitute them into (65), and then substitute three terms of each expansion into (66-68), we will obtain the following:

$$I_9 \approx \frac{2\pi}{\sqrt{R^2 + z_d^2 - 2zz_d + \rho^2 + 2\rho\sqrt{x^2+y^2}}}\left(1 + \frac{\rho\sqrt{x^2+y^2}}{R^2 + z_d^2 - 2zz_d + \rho^2 + 2\rho\sqrt{x^2+y^2}}\right).$$

$$I_{10} \approx \frac{\pi\left(x^2+y^2\right)\left(2R^2 + 2z_d^2 - 4zz_d + 2\rho^2 - 5\rho\sqrt{x^2+y^2}\right)}{2\left(R^2 + z_d^2 - 2zz_d + \rho^2 + 2\rho\sqrt{x^2+y^2}\right)^{5/2}}.$$

$$I_{11} \approx \frac{\pi\rho x\left(2R^2 + 2z_d^2 - 4zz_d + 2\rho^2 - 5\rho\sqrt{x^2+y^2}\right)}{2\left(R^2 + z_d^2 - 2zz_d + \rho^2 + 2\rho\sqrt{x^2+y^2}\right)^{5/2}}.$$

$$I_{12} \approx \frac{\pi\rho y\left(2R^2 + 2z_d^2 - 4zz_d + 2\rho^2 - 5\rho\sqrt{x^2+y^2}\right)}{2\left(R^2 + z_d^2 - 2zz_d + \rho^2 + 2\rho\sqrt{x^2+y^2}\right)^{5/2}}. \qquad (70)$$

In (70) the quantities $I_{10}$, $I_{11}$ and $I_{12}$ are proportional to each other, so that their substitution into the expression for the potential leads to cancellation of the terms:

$$\varphi \approx \frac{\rho_{0q}\gamma_c}{4\pi\varepsilon_0}\int_{-a}^{a}\int_{0}^{\rho_d}\left\{\left[1 - \frac{2\pi\eta\rho_0\left(\rho^2 + z_d^2\right)}{3c^2}\right]I_9 + \frac{\omega^2\rho^2}{2c^2}I_{10}\right\}\rho\, d\rho\, dz_d .$$

Taking into account (70) and the relation $\rho_d = \sqrt{a^2 - z_d^2}$, we will calculate the integrals $\int_0^{\rho_d} I_9\, \rho\, d\rho$, $\int_0^{\rho_d} I_9\, \rho^3\, d\rho$ and $\int_0^{\rho_d} I_{10}\, \rho^3\, d\rho$:

$$H_1 = \int_0^{\rho_d} I_9 \,\rho\, d\rho = 2\pi\sqrt{R^2 + a^2 - 2zz_d + 2\sqrt{a^2 - z_d^2}\sqrt{x^2 + y^2}} -$$

$$-2\pi\sqrt{R^2 + z_d^2 - 2zz_d} + \frac{2\pi\left(x^2 + y^2\right)^2}{\left(z - z_d\right)^2\sqrt{R^2 + a^2 - 2zz_d + 2\sqrt{a^2 - z_d^2}\sqrt{x^2 + y^2}}} -$$

$$-\frac{2\pi\left(x^2 + y^2\right)\sqrt{R^2 + z_d^2 - 2zz_d}}{\left(z - z_d\right)^2} + \frac{2\pi\left(x^2 + y^2\right)^{3/2}\sqrt{a^2 - z_d^2}}{\left(z - z_d\right)^2\sqrt{R^2 + a^2 - 2zz_d + 2\sqrt{a^2 - z_d^2}\sqrt{x^2 + y^2}}} +$$

$$+\frac{2\pi\left(x^2 + y^2\right)}{\sqrt{R^2 + a^2 - 2zz_d + 2\sqrt{a^2 - z_d^2}\sqrt{x^2 + y^2}}} - \frac{2\pi\sqrt{x^2 + y^2}\sqrt{a^2 - z_d^2}}{\sqrt{R^2 + a^2 - 2zz_d + 2\sqrt{a^2 - z_d^2}\sqrt{x^2 + y^2}}}.$$

$$H_2 = \int_0^{\rho_d} I_9 \,\rho^3\, d\rho = \frac{2\pi}{3}\left(a^2 - 2R^2 - 3z_d^2 + 4zz_d\right)\sqrt{R^2 + a^2 - 2zz_d + 2\sqrt{a^2 - z_d^2}\sqrt{x^2 + y^2}} -$$

$$-\frac{2\pi\left(x^2 + y^2\right)^2}{\left(z - z_d\right)^2}\sqrt{R^2 + z_d^2 - 2zz_d} + 10\pi\left(x^2 + y^2\right)^{3/2} \operatorname{Arsh}\frac{\sqrt{a^2 - z_d^2} + \sqrt{x^2 + y^2}}{z - z_d} +$$

$$+\frac{2\pi\sqrt{a^2 - z_d^2}\sqrt{x^2 + y^2}\left(2R^2 + a^2 + z_d^2 - 4zz_d + 5\sqrt{a^2 - z_d^2}\sqrt{x^2 + y^2}\right)}{\sqrt{R^2 + a^2 - 2zz_d + 2\sqrt{a^2 - z_d^2}\sqrt{x^2 + y^2}}} -$$

$$-\frac{2\pi\sqrt{a^2 - z_d^2}\left(x^2 + y^2\right)^{3/2}\left(R^2 + z_d^2 - 2zz_d\right)}{\left(z - z_d\right)^2\sqrt{R^2 + a^2 - 2zz_d + 2\sqrt{a^2 - z_d^2}\sqrt{x^2 + y^2}}} - 10\pi\left(x^2 + y^2\right)^{3/2} \operatorname{Arsh}\frac{\sqrt{x^2 + y^2}}{z - z_d} -$$

$$-\frac{2\pi\left(a^2 - z_d^2\right)\left(x^2 + y^2\right)^2}{\left(z - z_d\right)^2\sqrt{R^2 + a^2 - 2zz_d + 2\sqrt{a^2 - z_d^2}\sqrt{x^2 + y^2}}} + 8\pi\left(x^2 + y^2\right)\sqrt{R^2 + z_d^2 - 2zz_d} +$$

$$+\frac{2\pi\left(x^2 + y^2\right)^2}{\left(z - z_d\right)^2}\sqrt{R^2 + a^2 - 2zz_d + 2\sqrt{a^2 - z_d^2}\sqrt{x^2 + y^2}} + \frac{4\pi}{3}\left(R^2 + z_d^2 - 2zz_d\right)^{3/2} -$$

$$-\frac{8\pi}{3}\left[\sqrt{a^2 - z_d^2}\sqrt{x^2 + y^2} + 3\left(x^2 + y^2\right)\right]\sqrt{R^2 + a^2 - 2zz_d + 2\sqrt{a^2 - z_d^2}\sqrt{x^2 + y^2}}.$$

$$H_3 = \int_0^{\rho_d} I_{10}\, \rho^3\, d\rho =$$

$$= \pi\left(x^2+y^2\right)^{3/2} \left[ \begin{array}{l}
\dfrac{2R^2+a^2+z_d^2-4zz_d+5\sqrt{x^2+y^2}\sqrt{a^2-z_d^2}}{\sqrt{x^2+y^2}\sqrt{R^2+a^2-2zz_d+2\sqrt{x^2+y^2}\sqrt{a^2-z_d^2}}} - \dfrac{2\sqrt{R^2+z_d^2-2zz_d}}{\sqrt{x^2+y^2}} - \\
-\dfrac{\sqrt{x^2+y^2}\left(R^2+z_d^2-2zz_d\right)+\left(x^2+y^2\right)\sqrt{a^2-z_d^2}}{\left(z-z_d\right)^2\sqrt{R^2+a^2-2zz_d+2\sqrt{x^2+y^2}\sqrt{a^2-z_d^2}}} - \\
-\dfrac{3\left(a^2-z_d^2\right)^2\left(\sqrt{a^2-z_d^2}+\sqrt{x^2+y^2}\right)}{2\left(z-z_d\right)^2\left(R^2+a^2-2zz_d+2\sqrt{a^2-z_d^2}\sqrt{x^2+y^2}\right)^{3/2}} - \\
-\dfrac{6\left(a^2-z_d^2\right)^{3/2}}{\left(z-z_d\right)^2\sqrt{R^2+a^2-2zz_d+2\sqrt{a^2-z_d^2}\sqrt{x^2+y^2}}} - \\
-\dfrac{3\left(a^2-z_d^2\right)^2\left(\sqrt{a^2-z_d^2}+\sqrt{x^2+y^2}\right)}{\left(z-z_d\right)^4\sqrt{R^2+a^2-2zz_d+2\sqrt{a^2-z_d^2}\sqrt{x^2+y^2}}} - \\
-\dfrac{15\sqrt{x^2+y^2}-9\sqrt{a^2-z_d^2}}{2\left(z-z_d\right)^2}\sqrt{R^2+a^2-2zz_d+2\sqrt{a^2-z_d^2}\sqrt{x^2+y^2}} + \\
+\dfrac{17\sqrt{x^2+y^2}}{2\left(z-z_d\right)^2}\sqrt{R^2+z_d^2-2zz_d} + \dfrac{15}{2}\mathrm{Arsh}\dfrac{\sqrt{x^2+y^2}}{z-z_d} + \\
+\dfrac{3\left(a^2-z_d^2\right)^{3/2}}{\left(z-z_d\right)^4}\sqrt{R^2+a^2-2zz_d+2\sqrt{a^2-z_d^2}\sqrt{x^2+y^2}} - \\
-\dfrac{3\sqrt{x^2+y^2}\left(a^2-z_d^2\right)}{\left(z-z_d\right)^4}\sqrt{R^2+a^2-2zz_d+2\sqrt{a^2-z_d^2}\sqrt{x^2+y^2}} + \\
+\dfrac{3\left(x^2+y^2\right)\sqrt{a^2-z_d^2}}{\left(z-z_d\right)^4}\sqrt{R^2+a^2-2zz_d+2\sqrt{a^2-z_d^2}\sqrt{x^2+y^2}} - \\
-\dfrac{3\left(x^2+y^2\right)^{3/2}}{\left(z-z_d\right)^4}\sqrt{R^2+a^2-2zz_d+2\sqrt{a^2-z_d^2}\sqrt{x^2+y^2}} + \\
+\dfrac{3\left(x^2+y^2\right)^{3/2}}{\left(z-z_d\right)^4}\sqrt{R^2+z_d^2-2zz_d} - \dfrac{15}{2}\mathrm{Arsh}\dfrac{\sqrt{a^2-z_d^2}+\sqrt{x^2+y^2}}{z-z_d}
\end{array} \right]. \tag{71}$$

Now the potential can be expressed in terms of the integrals over the variable $z_d$ of the above quantities, $H_1$, $H_2$ and $H_3$:

$$\varphi \approx \frac{\rho_{0q}\gamma_c}{4\pi\varepsilon_0}\int_{-a}^{a} H_1\,dz_d + \frac{\omega^2\rho_{0q}\gamma_c}{8\pi c^2\varepsilon_0}\int_{-a}^{a} H_3\,dz_d - \frac{\eta\,\rho_0\,\rho_{0q}\gamma_c}{6c^2\varepsilon_0}\int_{-a}^{a} z_d^2 H_1\,dz_d - \frac{\eta\,\rho_0\rho_{0q}\gamma_c}{6c^2\varepsilon_0}\int_{-a}^{a} H_2\,dz_d \;. \tag{72}$$

Due to the cumbersomeness of the expressions for $H_1$, $H_2$ and $H_3$, integration in (72) becomes difficult, besides the solution is expressed in terms of special functions and cannot be represented in an explicit form without expansion into series. In this regard, we will consider here only three simplest cases.

The first term on the right-hand side of (72), that is, the term $\varphi' = \frac{\rho_{0q}\gamma_c}{4\pi\varepsilon_0}\int_{-a}^{a} H_1\,dz_d$, does not contain the speed of light and does not depend on the angular velocity of rotation $\omega$. In the case of a classical uniform solid body and in the absence of rotation, this term should define the scalar potential in accordance with Coulomb law. Indeed, if we calculate $\varphi'$ using $H_1$ from (71) on the axis $OZ$, on condition that $x = y = 0$, $z = R$, then we will obtain:

$$H_1(z=R) = 2\pi\sqrt{R^2 + a^2 - 2Rz_d} - 2\pi(R - z_d).$$

$$\varphi'(z=R) = \frac{\rho_{0q}\gamma_c}{4\pi\varepsilon_0}\int_{-a}^{a} H_1(z=R)\,dz_d = \frac{\rho_{0q}a^3\gamma_c}{3\varepsilon_0 R}\,. \tag{73}$$

In a solid body the Lorentz factor at the center of the sphere is $\gamma_c = 1$. Taking into account that the electric charge of a uniformly charged solid spherical body is $q = \frac{4\pi\rho_{0q}a^3}{3}$, we have: $\varphi'(z=R) = \frac{q}{4\pi\varepsilon_0 R}$, which corresponds to the Coulomb law on the axis $OZ$.

In the case of a relativistic uniform system, potential (72) on the axis $OZ$, on condition that $x = y = 0$, $z = R$, will depend only on $H_1$ and $H_2$, since $H_3$ vanishes. Since

$$H_2(z=R) = \frac{2\pi}{3}\left(a^2 - 2R^2 - 3z_d^2 + 4Rz_d\right)\sqrt{R^2 + a^2 - 2Rz_d} + \frac{4\pi}{3}(R - z_d)^3,$$

then, in view of (73) and (31), potential (72) becomes equal to:

$$\varphi(z=R)\approx\frac{\rho_{0q}a^3\gamma_c}{3\varepsilon_0 R}\left(1-\frac{2\pi\eta\,\rho_0\,a^2}{5c^2}\right)\approx\frac{q_\omega}{4\pi\varepsilon_0 R}. \tag{74}$$

Determination of the potential on the sphere's surface, where $z=0$, $R=a$, is of particular interest. Using $H_1$ from (71), we will express $\int_{-a}^{a} H_1\,dz_d$ in (72) in the following form:

$$\begin{aligned}\int_{-a}^{a} H_1\,dz_d &= 2\pi I_{13}-2\pi\int_{-a}^{a}\sqrt{R^2+z_d^2-2zz_d}\,dz_d+2\pi I_{14}-\\ &-2\pi\left(x^2+y^2\right)\int_{-a}^{a}\frac{\sqrt{R^2+z_d^2-2zz_d}}{\left(z-z_d\right)^2}\,dz_d+2\pi I_{15}+2\pi I_{16}-2\pi I_{17}.\end{aligned} \tag{75}$$

Here

$$I_{13}=\int_{-a}^{a}\sqrt{R^2+a^2-2zz_d+2\sqrt{a^2-z_d^2}\sqrt{x^2+y^2}}\,dz_d\ .$$

$$I_{14}=\left(x^2+y^2\right)^2\int_{-a}^{a}\frac{dz_d}{\left(z-z_d\right)^2\sqrt{R^2+a^2-2zz_d+2\sqrt{a^2-z_d^2}\sqrt{x^2+y^2}}}\ .$$

$$I_{15}=\left(x^2+y^2\right)^{3/2}\int_{-a}^{a}\frac{\sqrt{a^2-z_d^2}}{\left(z-z_d\right)^2\sqrt{R^2+a^2-2zz_d+2\sqrt{a^2-z_d^2}\sqrt{x^2+y^2}}}\,dz_d\ .$$

$$I_{16}=\left(x^2+y^2\right)\int_{-a}^{a}\frac{1}{\sqrt{R^2+a^2-2zz_d+2\sqrt{a^2-z_d^2}\sqrt{x^2+y^2}}}\,dz_d\ .$$

$$I_{17}=\sqrt{x^2+y^2}\int_{-a}^{a}\frac{\sqrt{a^2-z_d^2}}{\sqrt{R^2+a^2-2zz_d+2\sqrt{a^2-z_d^2}\sqrt{x^2+y^2}}}\,dz_d\ .$$

When $z=0$, $R=a$, all the integrals in (75) are taken exactly, without applying elliptic integrals, by using substitution $z_d = a\sin 2\gamma$. In particular, we have:

$$I_{13}\left(R=a\right)=\frac{8\sqrt{2}a^2}{3}. \qquad I_{14}\left(R=a\right)=-\frac{a^2}{4}\ln\operatorname{tg}\left(\frac{\pi}{8}\right)-\frac{3\sqrt{2}a^2}{4}.$$

$$I_{15}\left(R=a\right)=-\frac{\sqrt{2}a^2}{4}+\frac{5a^2}{4}\ln\operatorname{tg}\left(\frac{\pi}{8}\right). \qquad I_{16}\left(R=a\right)=2\sqrt{2}a^2+2a^2\ln\operatorname{tg}\left(\frac{\pi}{8}\right).$$

$$I_{17}\left(R=a\right)=\frac{2\sqrt{2}a^2}{3}+2a^2\ln\operatorname{tg}\left(\frac{\pi}{8}\right). \tag{76}$$

Substituting (76) into (75), we find:

$$\int_{-a}^{a} H_1\left(R=a\right)dz_d = 2\pi\left[\frac{16\sqrt{2}a^2}{3}+5a^2\ln\operatorname{tg}\left(\frac{\pi}{8}\right)-3a^2\ln\left(1+\sqrt{2}\right)\right]\approx 0{,}98\pi a^2 \cdot$$

On the other hand, for the Coulomb law to hold true for a fixed solid body at $z=0$, $R=a$, in (72) only the first term is taken into account and it must be $\int_{-a}^{a} H_1\, dz_d = \frac{4\pi a^2}{3}\approx 1{,}32\,\pi a^2$. The obtained above value $0{,}98\pi a^2$ turns out to be 26% less. The difference arose from the fact that when calculating the integrals $I_9$ in (65) and $I_{10}$ in (66), expansion (69) of complete elliptic integrals was used only up to the second- and third-order terms, respectively. For greater accuracy, an increased number of expansion terms should be used.

Thus, it can be stated that the scalar potential outside the sphere is determined exactly on the axis $OZ$, and in the other directions we obtain only an approximate estimate, depending on the number of used expansion terms in (69). Nevertheless, since $H_1$ does not depend on either the speed of light or the angular velocity of rotation $\omega$, this also applies to the potential $\varphi' = \frac{\rho_{0q}\,\gamma_c}{4\pi\varepsilon_0}\int_{-a}^{a} H_1\, dz_d$ in (72). This means that the value of the potential $\varphi'$ in an arbitrary direction cannot differ significantly from the value $\varphi'\left(z=R\right)$ in (73) on the axis $OZ$ and from $\varphi\left(z=R\right)$ in (74). Indeed, the dependence of the potential on the direction of the radius-vector $\mathbf{R}$ from the center of the sphere to the point with coordinates $x, y, z$, where the potential is calculated, could arise due to rotation. However, the potential $\varphi'$ does not depend on $\omega$, and for the sphere at rest with $\omega = 0$ the potential is symmetric with respect to the choice of direction of the vector $\mathbf{R}$.

In this regard, we will assume that in (72)

$$\varphi' = \frac{\rho_{0q}\gamma_c}{4\pi\varepsilon_0}\int_{-a}^{a} H_1\, dz_d \approx \varphi'(z = R) = \frac{\rho_{0q} a^3 \gamma_c}{3\varepsilon_0 R}. \tag{77}$$

Now it remains for us to calculate the last three terms on the right-hand side of (72) using (71). The integrals $\int_{-a}^{a} z_d^2 H_1\, dz_d$ and $\int_{-a}^{a} H_2\, dz_d$ can be evaluated by their value at $z = R$:

$$\int_{-a}^{a} z_d^2 H_1(z = R)\, dz_d \approx 2\pi\left(\frac{2a^5}{15R} + \frac{4a^7}{105R^3}\ldots\right), \qquad \int_{-a}^{a} H_2(z = R)\, dz_d \approx 2\pi\left(\frac{4a^5}{15R} - \frac{4a^7}{105R^3}\ldots\right).$$

At small $z$, the second and subsequent terms in the brackets on the right-hand side of these expressions become different, but the first terms do not change. Therefore, in a first approximation, we can assume that

$$\int_{-a}^{a} z_d^2 H_1\, dz_d \approx \frac{4\pi a^5}{15R}, \qquad \int_{-a}^{a} H_2\, dz_d \approx \frac{8\pi a^5}{15R}. \tag{78}$$

As for the integral $\int_{-a}^{a} H_3\, dz_d$, it is equal to zero at $z = R$, $x^2 + y^2 = 0$, and increases to a maximum at $z = 0$.

From (71) and (70), after expanding $\left(R^2 + z_d^2 - 2zz_d + \rho^2 + 2\rho\sqrt{x^2 + y^2}\right)^{-5/2}$ into a series at large $R$, taking into account the relation $\rho_d^2 = a^2 - z_d^2$, we find approximately the following:

$$H_3(R>>a)=\int_0^{\rho_d} I_{10}\,\rho^3\,d\rho \approx \int_0^{\rho_d} \frac{\pi\left(x^2+y^2\right)\left(2R^2+2z_d^2-4zz_d+2\rho^2-5\rho\sqrt{x^2+y^2}\right)}{2\left(R^2+z_d^2-2zz_d+\rho^2+2\rho\sqrt{x^2+y^2}\right)^{5/2}}\rho^3\,d\rho \approx$$

$$\approx \frac{\pi\left(x^2+y^2\right)\left(a^2-z_d^2\right)^2}{4\left(R^2+z_d^2-2zz_d\right)^{3/2}}-\frac{\pi\left(x^2+y^2\right)}{2\left(R^2+z_d^2-2zz_d\right)^{5/2}}\left[\frac{\left(a^2-z_d^2\right)^3}{2}+3\left(a^2-z_d^2\right)^{5/2}\sqrt{x^2+y^2}\right].$$

Expanding again $\left(R^2+z_d^2-2zz_d\right)^{-3/2}$ and $\left(R^2+z_d^2-2zz_d\right)^{-5/2}$ into series at large $R$ up to the second- and first-order terms, respectively, we have:

$$\int_{-a}^{a} H_3(R>>a)\,dz_d \approx \frac{4\pi\left(x^2+y^2\right)a^5}{15R^3}-\frac{15\pi^2\left(x^2+y^2\right)^{3/2}a^6}{32R^5}-\frac{2\pi\left(x^2+y^2\right)a^7}{7R^5}.$$

With this in mind, it follows from (72), (77), (78) and (31) that the potential at rather large $R$ is equal to:

$$\varphi(R>>a)\approx \frac{q_\omega}{4\pi\varepsilon_0 R}\left[1+\frac{\omega^2a^2(x^2+y^2)}{10c^2R^2}\left(1-\frac{225\pi a\sqrt{x^2+y^2}}{128R^2}-\frac{15a^2}{14R^2}\right)\right]. \qquad (79)$$

Potential (79) actually has the same dependence on the angular velocity $\omega$ as potential (32) in the middle zone, but it is not exact in the near zone, where the radius $R$ is not much larger than the sphere’s radius $a$.

We can also estimate the potential in the case, when $z=0$, $R=a$, and all the integrals are taken quite easily. Using $H_3$ from (71), we find:

$$\int_{-a}^{a} H_3(R=a)\,dz_d = -\frac{1279\sqrt{2}\pi a^4}{240}-\frac{427\pi a^4}{16}\ln \operatorname{tg}\left(\frac{\pi}{8}\right)\approx 15{,}98\pi a^4 .$$

Instead of (79) for the potential we obtain the following:

$$\varphi(R=a)\approx \frac{q_\omega}{4\pi\varepsilon_0 a}\left(1+\frac{6\omega^2a^2}{c^2}\right). \qquad (80)$$

Comparison of (80) with (79) shows that in our calculations at $z=0$ on the surface of a rotating sphere, the correction with respect to the potential of a fixed sphere reaches the value of the order of $\frac{6\omega^2 a^2}{c^2}$.

### 2.9. Vector potential in the near zone

Based on the similarity of formulas for scalar potential (16) and vector potential (18), in view of (63) we can express the components of the vector potential of the rotating disk in the near zone:

$$A_{dx} \approx -\frac{\mu_0 \omega s \rho_{0q}}{4\pi} \int_0^{2\pi} \int_0^{\rho_d} \left[ \begin{array}{l} 1 - \frac{\omega^2 \rho^2 (x\sin\phi - y\cos\phi)^2}{2c^2 R_P^2} + \\ + \frac{\omega^2 \rho x}{c^2} \cos\phi + \frac{\omega^2 \rho y}{c^2} \sin\phi \end{array} \right] \frac{\gamma' \sin\hat{\phi}\, \rho^2 d\rho\, d\phi}{R_P}.$$

$$A_{dy} \approx \frac{\mu_0 \omega s \rho_{0q}}{4\pi} \int_0^{2\pi} \int_0^{\rho_d} \left[ \begin{array}{l} 1 - \frac{\omega^2 \rho^2 (x\sin\phi - y\cos\phi)^2}{2c^2 R_P^2} + \\ + \frac{\omega^2 \rho x}{c^2} \cos\phi + \frac{\omega^2 \rho y}{c^2} \sin\phi \end{array} \right] \frac{\gamma' \cos\hat{\phi}\, \rho^2 d\rho\, d\phi}{R_P}.$$

Assuming $\hat{R}_P \approx R_P$, instead of (19) we have the following:

$$\cos\hat{\phi} \approx \cos\phi + \frac{\omega R_P}{c} \sin\phi, \qquad \sin\hat{\phi} \approx \sin\phi - \frac{\omega R_P}{c} \cos\phi.$$

Taking this into account, we will transform the vector potential components $A_{dx}$ and $A_{dy}$:

$$A_{dx} \approx -\frac{\mu_0 \omega s \rho_{0q}}{4\pi} \int_0^{2\pi} \int_0^{\rho_d} \left[ \begin{array}{l} 1 - \frac{\omega^2 \rho^2 (x\sin\phi - y\cos\phi)^2}{2c^2 R_P^2} + \\ + \frac{\omega^2 \rho x}{c^2} \cos\phi + \frac{\omega^2 \rho y}{c^2} \sin\phi \end{array} \right] \frac{\gamma' \sin\phi\, \rho^2 d\rho\, d\phi}{R_P} +$$
$$+ \frac{\mu_0 \omega^2 s \rho_{0q}}{4\pi c} \int_0^{2\pi} \int_0^{\rho_d} \left[ \begin{array}{l} 1 - \frac{\omega^2 \rho^2 (x\sin\phi - y\cos\phi)^2}{2c^2 R_P^2} + \\ + \frac{\omega^2 \rho x}{c^2} \cos\phi + \frac{\omega^2 \rho y}{c^2} \sin\phi \end{array} \right] \gamma' \cos\phi\, \rho^2 d\rho\, d\phi.$$

$$A_{dy} \approx \frac{\mu_0 \omega s \rho_{0q}}{4\pi} \int_0^{2\pi} \int_0^{\rho_d} \left[ \begin{array}{l} 1 - \dfrac{\omega^2 \rho^2 (x \sin\phi - y \cos\phi)^2}{2c^2 R_P^2} + \\ + \dfrac{\omega^2 \rho x}{c^2} \cos\phi + \dfrac{\omega^2 \rho y}{c^2} \sin\phi \end{array} \right] \frac{\gamma' \cos\phi \, \rho^2 \, d\rho \, d\phi}{R_P} +$$
$$+ \frac{\mu_0 \omega^2 s \rho_{0q}}{4\pi c} \int_0^{2\pi} \int_0^{\rho_d} \left[ \begin{array}{l} 1 - \dfrac{\omega^2 \rho^2 (x \sin\phi - y \cos\phi)^2}{2c^2 R_P^2} + \\ + \dfrac{\omega^2 \rho x}{c^2} \cos\phi + \dfrac{\omega^2 \rho y}{c^2} \sin\phi \end{array} \right] \gamma' \sin\phi \, \rho^2 \, d\rho \, d\phi. \tag{81}$$

Here $R_P$ is defined in (62).

In addition to the integrals $I_{11}$ and $I_{12}$ from (64), when integrating over the angle in (81) it is necessary to calculate the following integrals:

$$I_{18} = \int_0^{2\pi} \frac{(x \sin\phi - y \cos\phi)^2 \sin\phi}{\left(R^2 + z_d^2 - 2zz_d + \rho^2 - 2\rho x \cos\phi - 2\rho y \sin\phi\right)^{3/2}} d\phi .$$

$$I_{19} = \int_0^{2\pi} \frac{(x \sin\phi - y \cos\phi)^2 \cos\phi}{\left(R^2 + z_d^2 - 2zz_d + \rho^2 - 2\rho x \cos\phi - 2\rho y \sin\phi\right)^{3/2}} d\phi .$$

$$I_{20} = \int_0^{2\pi} \frac{(x \sin\phi - y \cos\phi)^2 \sin\phi}{R^2 + z_d^2 - 2zz_d + \rho^2 - 2\rho x \cos\phi - 2\rho y \sin\phi} d\phi .$$

$$I_{21} = \int_0^{2\pi} \frac{(x \sin\phi - y \cos\phi)^2 \cos\phi}{R^2 + z_d^2 - 2zz_d + \rho^2 - 2\rho x \cos\phi - 2\rho y \sin\phi} d\phi .$$

$$I_{22} = \int_0^{2\pi} \frac{\sin^2\phi}{\sqrt{R^2 + z_d^2 - 2zz_d + \rho^2 - 2\rho x \cos\phi - 2\rho y \sin\phi}} d\phi .$$

$$I_{23} = \int_0^{2\pi} \frac{\cos^2\phi}{\sqrt{R^2 + z_d^2 - 2zz_d + \rho^2 - 2\rho x \cos\phi - 2\rho y \sin\phi}} d\phi .$$

$$I_{24} = \int_0^{2\pi} \frac{\sin\phi\cos\phi}{\sqrt{R^2 + z_d^2 - 2zz_d + \rho^2 - 2\rho x\cos\phi - 2\rho y\sin\phi}} d\phi .$$

(82)

Taking these integrals into account, (81) can be rewritten as follows:

$$A_{dx} \approx -\frac{\mu_0 \omega s \rho_{0q}}{4\pi} \int_0^{\rho_d} \left( I_{12} - \frac{\omega^2 \rho^2}{2c^2} I_{18} + \frac{\omega^2 \rho x}{c^2} I_{24} + \frac{\omega^2 \rho y}{c^2} I_{22} \right) \gamma' \rho^2 d\rho +$$
$$+ \frac{\mu_0 \omega^2 s \rho_{0q}}{4\pi c} \int_0^{\rho_d} \left( \frac{\pi \omega^2 \rho x}{c^2} - \frac{\omega^2 \rho^2}{2c^2} I_{21} \right) \gamma' \rho^2 d\rho .$$

$$A_{dy} \approx \frac{\mu_0 \omega s \rho_{0q}}{4\pi} \int_0^{\rho_d} \left( I_{11} - \frac{\omega^2 \rho^2}{2c^2} I_{19} + \frac{\omega^2 \rho x}{c^2} I_{23} + \frac{\omega^2 \rho y}{c^2} I_{24} \right) \gamma' \rho^2 d\rho +$$
$$+ \frac{\mu_0 \omega^2 s \rho_{0q}}{4\pi c} \int_0^{\rho_d} \left( \frac{\pi \omega^2 \rho y}{c^2} - \frac{\omega^2 \rho^2}{2c^2} I_{20} \right) \gamma' \rho^2 d\rho.$$

(83)

Integrals (82) are obtained in the following form:

$$I_{18} \approx -\frac{2\pi y \left( R^2 + z_d^2 - 2zz_d + \rho^2 + \rho\sqrt{x^2 + y^2} \right)}{\rho \left( R^2 + z_d^2 - 2zz_d + \rho^2 + 2\rho\sqrt{x^2 + y^2} \right)^{3/2}} .$$

$$I_{19} \approx -\frac{2\pi x \left( R^2 + z_d^2 - 2zz_d + \rho^2 + \rho\sqrt{x^2 + y^2} \right)}{\rho \left( R^2 + z_d^2 - 2zz_d + \rho^2 + 2\rho\sqrt{x^2 + y^2} \right)^{3/2}} .$$

$$I_{20} = \frac{\pi y \left( R^2 + z_d^2 - 2zz_d + \rho^2 \right)^2}{4\rho^3 \left( x^2 + y^2 \right)} - \frac{\pi y}{2\rho} .$$

$$I_{21} = \frac{\pi x \left( R^2 + z_d^2 - 2zz_d + \rho^2 \right)^2}{4\rho^3 \left( x^2 + y^2 \right)} - \frac{\pi x}{2\rho} .$$

$$I_{22} \approx \frac{2\pi x^2}{\left(x^2+y^2\right)\sqrt{R^2+z_d^2-2zz_d+\rho^2+2\rho\sqrt{x^2+y^2}}}+$$

$$+\frac{2\pi\rho y^2}{\sqrt{x^2+y^2}\left(R^2+z_d^2-2zz_d+\rho^2+2\rho\sqrt{x^2+y^2}\right)^{3/2}}.$$

$$I_{23} \approx \frac{2\pi y^2}{\left(x^2+y^2\right)\sqrt{R^2+z_d^2-2zz_d+\rho^2+2\rho\sqrt{x^2+y^2}}}+$$

$$+\frac{2\pi\rho x^2}{\sqrt{x^2+y^2}\left(R^2+z_d^2-2zz_d+\rho^2+2\rho\sqrt{x^2+y^2}\right)^{3/2}}.$$

$$I_{24} \approx -\frac{2\pi xy\left(R^2+z_d^2-2zz_d+\rho^2+\rho\sqrt{x^2+y^2}\right)}{\left(x^2+y^2\right)\left(R^2+z_d^2-2zz_d+\rho^2+2\rho\sqrt{x^2+y^2}\right)^{3/2}}.$$

When calculating $I_{18}$, $I_{19}$, $I_{22}$, $I_{23}$, $I_{24}$, elliptic integrals appear, which, using (69), have been expanded to the second-order terms.

Taking these integrals into account, as well as (27) and the integrals $I_{11}$, $I_{12}$ from (70), expressions (83) take the following form:

$$A_{dx} \approx -\frac{\mu_0\omega\rho_{0q}ys\gamma_c}{4}\int_0^{\rho_d}\frac{\left\{\begin{array}{l}1-\dfrac{2\pi\eta\rho_0\left(\rho^2+z_d^2\right)}{3c^2}+\\+\dfrac{\omega^2\left(R^2+z_d^2-2zz_d+\rho^2+3\rho\sqrt{x^2+y^2}\right)}{c^2}-\\-\dfrac{9\rho\sqrt{x^2+y^2}\left[1-\dfrac{2\pi\eta\rho_0\left(\rho^2+z_d^2\right)}{3c^2}\right]}{2\left(R^2+z_d^2-2zz_d+\rho^2+2\rho\sqrt{x^2+y^2}\right)}\end{array}\right\}\rho^3 d\rho}{\left(R^2+z_d^2-2zz_d+\rho^2+2\rho\sqrt{x^2+y^2}\right)^{3/2}}-$$

$$-\frac{\mu_0\omega^4\rho_{0q}xs\gamma_c}{16c^3}\int_0^{\rho_d}\left[\frac{\left(R^2+z_d^2-2zz_d+\rho^2\right)^2}{2\rho\left(x^2+y^2\right)}-5\rho\right]\rho^2 d\rho.$$

$$A_{dy} \approx \frac{\mu_0 \omega \rho_{0q} x s \gamma_c}{4} \int_0^{\rho_d} \frac{\left\{ \begin{array}{l} 1 - \dfrac{2\pi \eta \rho_0 \left(\rho^2 + z_d^2\right)}{3c^2} + \\ + \dfrac{\omega^2 \left(R^2 + z_d^2 - 2zz_d + \rho^2 + 3\rho\sqrt{x^2+y^2}\right)}{c^2} - \\ - \dfrac{9\rho\sqrt{x^2+y^2}\left[1 - \dfrac{2\pi \eta \rho_0 \left(\rho^2 + z_d^2\right)}{3c^2}\right]}{2\left(R^2 + z_d^2 - 2zz_d + \rho^2 + 2\rho\sqrt{x^2+y^2}\right)} \end{array} \right\} \rho^3 d\rho}{\left(R^2 + z_d^2 - 2zz_d + \rho^2 + 2\rho\sqrt{x^2+y^2}\right)^{3/2}} -$$

$$- \frac{\mu_0 \omega^4 \rho_{0q} y s \gamma_c}{16c^3} \int_0^{\rho_d} \left( \frac{\left(R^2 + z_d^2 - 2zz_d + \rho^2\right)^2}{2\rho\left(x^2+y^2\right)} - 5\rho \right) \rho^2 d\rho.$$

If we denote

$$I_{25} = \int_0^{\rho_d} \frac{\rho^3 d\rho}{\sqrt{R^2 + z_d^2 - 2zz_d + \rho^2 + 2\rho\sqrt{x^2+y^2}}},$$

$$I_{26} = \int_0^{\rho_d} \frac{\rho^3 d\rho}{\left(R^2 + z_d^2 - 2zz_d + \rho^2 + 2\rho\sqrt{x^2+y^2}\right)^{3/2}},$$

$$I_{27} = \int_0^{\rho_d} \frac{\rho^4 d\rho}{\left(R^2 + z_d^2 - 2zz_d + \rho^2 + 2\rho\sqrt{x^2+y^2}\right)^{3/2}},$$

$$I_{28} = \int_0^{\rho_d} \frac{\rho^5 d\rho}{\left(R^2 + z_d^2 - 2zz_d + \rho^2 + 2\rho\sqrt{x^2+y^2}\right)^{3/2}},$$

$$I_{29} = \int_0^{\rho_d} \frac{\rho^4 d\rho}{\left(R^2 + z_d^2 - 2zz_d + \rho^2 + 2\rho\sqrt{x^2+y^2}\right)^{5/2}},$$

$$I_{30} = \int_0^{\rho_d} \frac{\rho^6 \, d\rho}{\left(R^2 + z_d^2 - 2zz_d + \rho^2 + 2\rho\sqrt{x^2+y^2}\right)^{5/2}}, \tag{84}$$

then the vector potential components, in view of the relation $\rho_d^2 = a^2 - z_d^2$, will be expressed as follows:

$$A_{dx} \approx -\frac{\mu_0 \omega \rho_{0q} y s \gamma_c}{4} \begin{bmatrix} \left(1 - \frac{2\pi\eta\rho_0 z_d^2}{3c^2}\right) I_{26} - \frac{2\pi\eta\rho_0}{3c^2} I_{28} + \frac{\omega^2}{c^2} I_{25} + \\ + \frac{\omega^2\sqrt{x^2+y^2}}{c^2} I_{27} + \frac{3\pi\eta\rho_0\sqrt{x^2+y^2}}{c^2} I_{30} + \\ + \left(\frac{3\pi\eta\rho_0 z_d^2\sqrt{x^2+y^2}}{c^2} - \frac{9\sqrt{x^2+y^2}}{2}\right) I_{29} \end{bmatrix} -$$

$$-\frac{\mu_0 \omega^4 \rho_{0q} x s \gamma_c}{192 c^3 \left(x^2+y^2\right)} \left[\left(R^2 + a^2 - 2zz_d\right)^3 - \left(R^2 + z_d^2 - 2zz_d\right)^3 - 15\left(x^2+y^2\right)\left(a^2 - z_d^2\right)^2\right].$$

$$A_{dy} \approx \frac{\mu_0 \omega \rho_{0q} x s \gamma_c}{4} \begin{bmatrix} \left(1 - \frac{2\pi\eta\rho_0 z_d^2}{3c^2}\right) I_{26} - \frac{2\pi\eta\rho_0}{3c^2} I_{28} + \frac{\omega^2}{c^2} I_{25} + \\ + \frac{\omega^2\sqrt{x^2+y^2}}{c^2} I_{27} + \frac{3\pi\eta\rho_0\sqrt{x^2+y^2}}{c^2} I_{30} + \\ + \left(\frac{3\pi\eta\rho_0 z_d^2\sqrt{x^2+y^2}}{c^2} - \frac{9\sqrt{x^2+y^2}}{2}\right) I_{29} \end{bmatrix} -$$

$$-\frac{\mu_0 \omega^4 \rho_{0q} y s \gamma_c}{192 c^3 \left(x^2+y^2\right)} \left[\left(R^2 + a^2 - 2zz_d\right)^3 - \left(R^2 + z_d^2 - 2zz_d\right)^3 - 15\left(x^2+y^2\right)\left(a^2 - z_d^2\right)^2\right].$$

The components $A_{dx}$ and $A_{dy}$ represent the vector potential components, arising from rotation of one layer in the form of a disk inside the sphere. To pass on to the component $A_x$ of the vector potential from the entire sphere, it is necessary to replace $s$ with the differential $dz_d$, and the sum of the components $A_{dx}$ should be considered as an integral over all the sphere's layers, which depends on the variable $z_d$. The same follows for $A_{dy}$. This gives the following:

$$A_x \approx -\frac{\mu_0 \omega \rho_{0q} y \gamma_c}{4} \int_{-a}^{a} \begin{bmatrix} \left(1 - \frac{2\pi \eta \rho_0 z_d^2}{3c^2}\right) I_{26} - \frac{2\pi \eta \rho_0}{3c^2} I_{28} + \frac{\omega^2}{c^2} I_{25} + \\ + \frac{\omega^2 \sqrt{x^2+y^2}}{c^2} I_{27} + \frac{3\pi \eta \rho_0 \sqrt{x^2+y^2}}{c^2} I_{30} + \\ + \left( \frac{3\pi \eta \rho_0 z_d^2 \sqrt{x^2+y^2}}{c^2} - \frac{9\sqrt{x^2+y^2}}{2} \right) I_{29} \end{bmatrix} dz_d -$$

$$-\frac{\mu_0 \omega^4 \rho_{0q} x y \gamma_c}{192 c^3 \left(x^2+y^2\right)} \int_{-a}^{a} \left[ \left(R^2 + a^2 - 2zz_d\right)^3 - \left(R^2 + z_d^2 - 2zz_d\right)^3 - 15\left(x^2+y^2\right)\left(a^2 - z_d^2\right)^2 \right] dz_d .$$

$$A_y \approx \frac{\mu_0 \omega \rho_{0q} x \gamma_c}{4} \int_{-a}^{a} \begin{bmatrix} \left(1 - \frac{2\pi \eta \rho_0 z_d^2}{3c^2}\right) I_{26} - \frac{2\pi \eta \rho_0}{3c^2} I_{28} + \frac{\omega^2}{c^2} I_{25} + \\ + \frac{\omega^2 \sqrt{x^2+y^2}}{c^2} I_{27} + \frac{3\pi \eta \rho_0 \sqrt{x^2+y^2}}{c^2} I_{30} + \\ + \left( \frac{3\pi \eta \rho_0 z_d^2 \sqrt{x^2+y^2}}{c^2} - \frac{9\sqrt{x^2+y^2}}{2} \right) I_{29} \end{bmatrix} dz_d -$$

$$-\frac{\mu_0 \omega^4 \rho_{0q} y y \gamma_c}{192 c^3 \left(x^2+y^2\right)} \int_{-a}^{a} \left[ \left(R^2 + a^2 - 2zz_d\right)^3 - \left(R^2 + z_d^2 - 2zz_d\right)^3 - 15\left(x^2+y^2\right)\left(a^2 - z_d^2\right)^2 \right] dz_d .$$

(85)

The integrals in (84), taking into account the relations $R^2 = x^2 + y^2 + z^2$, $R^2 + z_d^2 - 2zz_d + \rho^2 + 2\rho\sqrt{x^2+y^2} = \left(z - z_d\right)^2 + \left(\rho + \sqrt{x^2+y^2}\right)^2$, $\rho_d^2 = a^2 - z_d^2$, are equal to:

$$I_{25} = \frac{2a^2 + \sqrt{a^2 - z_d^2}\sqrt{x^2+y^2} - 2z_d^2 + 5\left(x^2+y^2\right)}{2} \sqrt{R^2 + a^2 - 2zz_d + 2\sqrt{a^2 - z_d^2}\sqrt{x^2+y^2}} -$$

$$-\frac{2}{3}\left(R^2 + a^2 - 2zz_d + 2\sqrt{a^2 - z_d^2}\sqrt{x^2+y^2}\right)^{3/2} +$$

$$+\frac{4R^2 + 4z_d^2 - 8zz_d - 15\left(x^2+y^2\right)}{6} \sqrt{R^2 + z_d^2 - 2zz_d} +$$

$$+\frac{\sqrt{x^2+y^2}}{2}\left[3z_d^2 - 6zz_d + 3R^2 - 5\left(x^2+y^2\right)\right] \operatorname{Arsh} \frac{\sqrt{a^2 - z_d^2} + \sqrt{x^2+y^2}}{z - z_d} -$$

$$-\frac{\sqrt{x^2+y^2}}{2}\left[3z_d^2 - 6zz_d + 3R^2 - 5\left(x^2+y^2\right)\right] \operatorname{Arsh} \frac{\sqrt{x^2+y^2}}{z - z_d} .$$

$$I_{26}=\frac{2R^2+a^2+z_d^2-4zz_d+5\sqrt{x^2+y^2}\sqrt{a^2-z_d^2}}{\sqrt{R^2+a^2-2zz_d+2\sqrt{x^2+y^2}\sqrt{a^2-z_d^2}}}-2\sqrt{R^2+z_d^2-2zz_d}-$$

$$-\frac{\left(x^2+y^2\right)\left(R^2+z_d^2-2zz_d\right)+\left(x^2+y^2\right)^{3/2}\sqrt{a^2-z_d^2}}{\left(z-z_d\right)^2\sqrt{R^2+a^2-2zz_d+2\sqrt{x^2+y^2}\sqrt{a^2-z_d^2}}}+\frac{\left(x^2+y^2\right)\sqrt{R^2+z_d^2-2zz_d}}{\left(z-z_d\right)^2}-$$

$$-3\sqrt{x^2+y^2}\,\mathrm{Arsh}\frac{\sqrt{a^2-z_d^2}+\sqrt{x^2+y^2}}{z-z_d}+3\sqrt{x^2+y^2}\,\mathrm{Arsh}\frac{\sqrt{x^2+y^2}}{z-z_d}.$$

$$I_{27}=-\frac{\sqrt{a^2-z_d^2}\left(x^2+y^2\right)\left(R^2+z_d^2-2zz_d\right)+\left(a^2-z_d^2\right)\left(x^2+y^2\right)^{3/2}}{\left(z-z_d\right)^2\sqrt{R^2+a^2-2zz_d+2\sqrt{a^2-z_d^2}\sqrt{x^2+y^2}}}+$$

$$+\frac{\sqrt{a^2-z_d^2}\left(2R^2+a^2+z_d^2-4zz_d+5\sqrt{a^2-z_d^2}\sqrt{x^2+y^2}\right)}{\sqrt{R^2+a^2-2zz_d+2\sqrt{a^2-z_d^2}\sqrt{x^2+y^2}}}-\frac{15}{2}\left(x^2+y^2\right)\mathrm{Arsh}\frac{\sqrt{x^2+y^2}}{z-z_d}+$$

$$+\frac{\left(x^2+y^2\right)^{3/2}}{\left(z-z_d\right)^2}\sqrt{R^2+a^2-2zz_d+2\sqrt{a^2-z_d^2}\sqrt{x^2+y^2}}-\frac{\left(x^2+y^2\right)^{3/2}}{\left(z-z_d\right)^2}\sqrt{R^2+z_d^2-2zz_d}-$$

$$-\frac{\sqrt{a^2-z_d^2}+13\sqrt{x^2+y^2}}{2}\sqrt{R^2+a^2-2zz_d+2\sqrt{a^2-z_d^2}\sqrt{x^2+y^2}}+$$

$$+\frac{15}{2}\left(x^2+y^2\right)\mathrm{Arsh}\frac{\sqrt{a^2-z_d^2}+\sqrt{x^2+y^2}}{z-z_d}+\frac{13\sqrt{x^2+y^2}}{2}\sqrt{R^2+z_d^2-2zz_d}-$$

$$-\frac{3}{2}\left(R^2+z_d^2-2zz_d\right)\mathrm{Arsh}\frac{\sqrt{a^2-z_d^2}+\sqrt{x^2+y^2}}{z-z_d}+\frac{3}{2}\left(R^2+z_d^2-2zz_d\right)\mathrm{Arsh}\frac{\sqrt{x^2+y^2}}{z-z_d}.$$

$$I_{28} = -\frac{\left(a^2 - z_d^2\right)\left(x^2 + y^2\right)^2}{\left(z - z_d\right)^2 \sqrt{R^2 + a^2 - 2zz_d + 2\sqrt{a^2 - z_d^2}\sqrt{x^2 + y^2}}} -$$

$$-\frac{\left(a^2 - z_d^2\right)^{3/2}\left(x^2 + y^2\right)^{3/2}}{\left(z - z_d\right)^2 \sqrt{R^2 + a^2 - 2zz_d + 2\sqrt{a^2 - z_d^2}\sqrt{x^2 + y^2}}} +$$

$$+\frac{\sqrt{a^2 - z_d^2}\left(x^2 + y^2\right)^{3/2}}{\left(z - z_d\right)^2}\sqrt{R^2 + a^2 - 2zz_d + 2\sqrt{a^2 - z_d^2}\sqrt{x^2 + y^2}} -$$

$$-\frac{\left(x^2 + y^2\right)^2}{\left(z - z_d\right)^2}\sqrt{R^2 + a^2 - 2zz_d + 2\sqrt{a^2 - z_d^2}\sqrt{x^2 + y^2}} + \frac{\left(x^2 + y^2\right)^2}{\left(z - z_d\right)^2}\sqrt{R^2 + z_d^2 - 2zz_d} -$$

$$-\frac{\left(a^2 - z_d^2\right)\left(x^2 + y^2\right)}{\sqrt{R^2 + a^2 - 2zz_d + 2\sqrt{a^2 - z_d^2}\sqrt{x^2 + y^2}}} - \frac{\left(a^2 - z_d^2\right)^{3/2}\left(\sqrt{a^2 - z_d^2} - \sqrt{x^2 + y^2}\right)}{\sqrt{R^2 + a^2 - 2zz_d + 2\sqrt{a^2 - z_d^2}\sqrt{x^2 + y^2}}} +$$

$$+\frac{4\left(a^2 - z_d^2\right)}{3}\sqrt{R^2 + a^2 - 2zz_d + 2\sqrt{a^2 - z_d^2}\sqrt{x^2 + y^2}} +$$

$$+\frac{33\left(x^2 + y^2\right)}{2}\sqrt{R^2 + a^2 - 2zz_d + 2\sqrt{a^2 - z_d^2}\sqrt{x^2 + y^2}} - \frac{33\left(x^2 + y^2\right)}{2}\sqrt{R^2 + z_d^2 - 2zz_d} -$$

$$-\frac{29\sqrt{a^2 - z_d^2}\sqrt{x^2 + y^2}}{6}\sqrt{R^2 + a^2 - 2zz_d + 2\sqrt{a^2 - z_d^2}\sqrt{x^2 + y^2}} -$$

$$-\frac{8}{3}\left(R^2 + z_d^2 - 2zz_d\right)\sqrt{R^2 + a^2 - 2zz_d + 2\sqrt{a^2 - z_d^2}\sqrt{x^2 + y^2}} + \frac{8}{3}\left(R^2 + z_d^2 - 2zz_d\right)^{3/2} +$$

$$+\frac{15\sqrt{x^2 + y^2}\left(R^2 + z_d^2 - 2zz_d\right)}{2}\operatorname{Arsh}\frac{\sqrt{a^2 - z_d^2} + \sqrt{x^2 + y^2}}{z - z_d} -$$

$$-\frac{15\sqrt{x^2 + y^2}\left(R^2 + z_d^2 - 2zz_d\right)}{2}\operatorname{Arsh}\frac{\sqrt{x^2 + y^2}}{z - z_d} -$$

$$-\frac{35\left(x^2 + y^2\right)^{3/2}}{2}\operatorname{Arsh}\frac{\sqrt{a^2 - z_d^2} + \sqrt{x^2 + y^2}}{z - z_d} + \frac{35\left(x^2 + y^2\right)^{3/2}}{2}\operatorname{Arsh}\frac{\sqrt{x^2 + y^2}}{z - z_d}.$$

$$I_{29} = \frac{\left(a^2 - z_d^2\right)^2 \left(\sqrt{a^2 - z_d^2} + \sqrt{x^2 + y^2}\right)}{3\left(z - z_d\right)^2 \left(R^2 + a^2 - 2zz_d + 2\sqrt{a^2 - z_d^2}\sqrt{x^2 + y^2}\right)^{3/2}} +$$

$$+ \frac{4\left(a^2 - z_d^2\right)^{3/2}}{3\left(z - z_d\right)^2 \sqrt{R^2 + a^2 - 2zz_d + 2\sqrt{a^2 - z_d^2}\sqrt{x^2 + y^2}}} +$$

$$+ \frac{2\left(a^2 - z_d^2\right)^2 \left(\sqrt{a^2 - z_d^2} + \sqrt{x^2 + y^2}\right)}{3\left(z - z_d\right)^4 \sqrt{R^2 + a^2 - 2zz_d + 2\sqrt{a^2 - z_d^2}\sqrt{x^2 + y^2}}} +$$

$$+ \frac{5\sqrt{x^2 + y^2} - 3\sqrt{a^2 - z_d^2}}{3\left(z - z_d\right)^2} \sqrt{R^2 + a^2 - 2zz_d + 2\sqrt{a^2 - z_d^2}\sqrt{x^2 + y^2}} -$$

$$- \frac{5\sqrt{x^2 + y^2}}{3\left(z - z_d\right)^2} \sqrt{R^2 + z_d^2 - 2zz_d} - \frac{2\left(a^2 - z_d^2\right)^{3/2}}{3\left(z - z_d\right)^4} \sqrt{R^2 + a^2 - 2zz_d + 2\sqrt{a^2 - z_d^2}\sqrt{x^2 + y^2}} +$$

$$+ \frac{2\left(a^2 - z_d^2\right)\sqrt{x^2 + y^2}}{3\left(z - z_d\right)^4} \sqrt{R^2 + a^2 - 2zz_d + 2\sqrt{a^2 - z_d^2}\sqrt{x^2 + y^2}} -$$

$$- \frac{2\sqrt{a^2 - z_d^2}\left(x^2 + y^2\right)}{3\left(z - z_d\right)^4} \sqrt{R^2 + a^2 - 2zz_d + 2\sqrt{a^2 - z_d^2}\sqrt{x^2 + y^2}} +$$

$$+ \frac{2\left(x^2 + y^2\right)^{3/2}}{3\left(z - z_d\right)^4} \sqrt{R^2 + a^2 - 2zz_d + 2\sqrt{a^2 - z_d^2}\sqrt{x^2 + y^2}} -$$

$$- \frac{2\left(x^2 + y^2\right)^{3/2}}{3\left(z - z_d\right)^4} \sqrt{R^2 + z_d^2 - 2zz_d} + \operatorname{Arsh}\frac{\sqrt{a^2 - z_d^2} + \sqrt{x^2 + y^2}}{z - z_d} - \operatorname{Arsh}\frac{\sqrt{x^2 + y^2}}{z - z_d}.$$

$$I_{30} = \frac{\left(a^2 - z_d^2\right)^3 \left(\sqrt{a^2 - z_d^2} + \sqrt{x^2 + y^2}\right)}{3\left(z - z_d\right)^2 \left(R^2 + a^2 - 2zz_d + 2\sqrt{a^2 - z_d^2}\sqrt{x^2 + y^2}\right)^{3/2}} +$$

$$+\frac{2\left(a^2 - z_d^2\right)^{7/2}}{3\left(z - z_d\right)^4 \sqrt{R^2 + a^2 - 2zz_d + 2\sqrt{a^2 - z_d^2}\sqrt{x^2 + y^2}}} +$$

$$+\frac{2\left(a^2 - z_d^2\right)^3 \sqrt{x^2 + y^2}}{3\left(z - z_d\right)^4 \sqrt{R^2 + a^2 - 2zz_d + 2\sqrt{a^2 - z_d^2}\sqrt{x^2 + y^2}}} +$$

$$+\frac{2\left(a^2 - z_d^2\right)^2 \left(x^2 + y^2\right)^{3/2}}{3\left(z - z_d\right)^4 \sqrt{R^2 + a^2 - 2zz_d + 2\sqrt{a^2 - z_d^2}\sqrt{x^2 + y^2}}} +$$

$$+\frac{4\left(a^2 - z_d^2\right)^{3/2} \left(x^2 + y^2\right)^2}{3\left(z - z_d\right)^4 \sqrt{R^2 + a^2 - 2zz_d + 2\sqrt{a^2 - z_d^2}\sqrt{x^2 + y^2}}} +$$

$$+\frac{2\left(a^2 - z_d^2\right)\left(x^2 + y^2\right)^{5/2}}{3\left(z - z_d\right)^4 \sqrt{R^2 + a^2 - 2zz_d + 2\sqrt{a^2 - z_d^2}\sqrt{x^2 + y^2}}} +$$

$$+\frac{2\left(a^2 - z_d^2\right)^{5/2}}{\left(z - z_d\right)^2 \sqrt{R^2 + a^2 - 2zz_d + 2\sqrt{a^2 - z_d^2}\sqrt{x^2 + y^2}}} +$$

$$+\frac{2\left(a^2 - z_d^2\right)\left(x^2 + y^2\right)^{3/2}}{3\left(z - z_d\right)^2 \sqrt{R^2 + a^2 - 2zz_d + 2\sqrt{a^2 - z_d^2}\sqrt{x^2 + y^2}}} -$$

$$-\frac{2\left(a^2 - z_d^2\right)^{5/2}}{3\left(z - z_d\right)^4}\sqrt{R^2 + a^2 - 2zz_d + 2\sqrt{a^2 - z_d^2}\sqrt{x^2 + y^2}} +$$

$$+\frac{2\left(a^2 - z_d^2\right)^2 \sqrt{x^2 + y^2}}{3\left(z - z_d\right)^4}\sqrt{R^2 + a^2 - 2zz_d + 2\sqrt{a^2 - z_d^2}\sqrt{x^2 + y^2}} -$$

$$-\frac{2\left(a^2 - z_d^2\right)^{3/2}\left(x^2 + y^2\right)}{3\left(z - z_d\right)^4}\sqrt{R^2 + a^2 - 2zz_d + 2\sqrt{a^2 - z_d^2}\sqrt{x^2 + y^2}} -$$

$$-\frac{2\sqrt{a^2 - z_d^2}\left(x^2 + y^2\right)^2}{3\left(z - z_d\right)^4}\sqrt{R^2 + a^2 - 2zz_d + 2\sqrt{a^2 - z_d^2}\sqrt{x^2 + y^2}} +$$

$$+\frac{2\left(x^2+y^2\right)^{5/2}}{3\left(z-z_d\right)^4}\sqrt{R^2+a^2-2zz_d+2\sqrt{a^2-z_d^2}\sqrt{x^2+y^2}}-$$

$$-\frac{2\left(x^2+y^2\right)^{5/2}}{3\left(z-z_d\right)^4}\sqrt{R^2+z_d^2-2zz_d}-\frac{5\left(a^2-z_d^2\right)^{3/2}}{3\left(z-z_d\right)^2}\sqrt{R^2+a^2-2zz_d+2\sqrt{a^2-z_d^2}\sqrt{x^2+y^2}}+$$

$$+\frac{3\sqrt{x^2+y^2}\left(a^2-z_d^2\right)}{\left(z-z_d\right)^2}\sqrt{R^2+a^2-2zz_d+2\sqrt{a^2-z_d^2}\sqrt{x^2+y^2}}-$$

$$-\frac{4\sqrt{a^2-z_d^2}\left(x^2+y^2\right)}{\left(z-z_d\right)^2}\sqrt{R^2+a^2-2zz_d+2\sqrt{a^2-z_d^2}\sqrt{x^2+y^2}}+$$

$$+\frac{14\left(x^2+y^2\right)^{3/2}}{3\left(z-z_d\right)^2}\sqrt{R^2+a^2-2zz_d+2\sqrt{a^2-z_d^2}\sqrt{x^2+y^2}}-\frac{14\left(x^2+y^2\right)^{3/2}}{3\left(z-z_d\right)^2}\sqrt{R^2+z_d^2-2zz_d}-$$

$$-\frac{27\sqrt{x^2+y^2}}{2}\sqrt{R^2+a^2-2zz_d+2\sqrt{a^2-z_d^2}\sqrt{x^2+y^2}}+\frac{27\sqrt{x^2+y^2}}{2}\sqrt{R^2+z_d^2-2zz_d}+$$

$$+\frac{5\sqrt{a^2-z_d^2}}{2}\sqrt{R^2+a^2-2zz_d+2\sqrt{a^2-z_d^2}\sqrt{x^2+y^2}}-$$

$$-\frac{5\left(z-z_d\right)^2}{2}\operatorname{Arsh}\frac{\sqrt{a^2-z_d^2}+\sqrt{x^2+y^2}}{z-z_d}+\frac{5\left(z-z_d\right)^2}{2}\operatorname{Arsh}\frac{\sqrt{x^2+y^2}}{z-z_d}+$$

$$+15\left(x^2+y^2\right)\operatorname{Arsh}\frac{\sqrt{a^2-z_d^2}+\sqrt{x^2+y^2}}{z-z_d}-15\left(x^2+y^2\right)\operatorname{Arsh}\frac{\sqrt{x^2+y^2}}{z-z_d}. \tag{86}$$

Now we can use the above expressions for the integrals $I_{25}$, $I_{26}$, $I_{27}$, $I_{28}$, $I_{29}$ and $I_{30}$ to substitute them into (85) and calculate the vector potential components.

Here it should be recalled that in the course of calculations, the integrals $I_{11}$ and $I_{12}$, defined in (64), (67), and (68), were calculated in (70) approximately, by using in (69) expansion up to the second-order terms. The same holds true for the integrals $I_{18}$, $I_{19}$, $I_{22}$, $I_{23}$, $I_{24}$. All this led to the appearance of expressions (83), which in the general case are still insufficient for exact calculation of the potential components.

A similar situation took place in the previous section, where we found that deviation of the scalar potential in our calculations in the equatorial plane on the sphere's surface reached 26% due to the fact that not all the expansion terms were used in (69). Therefore, it should be expected that although the dependence of the vector potential components on the coordinates $x, y, z$ is shown correctly in (85), the inaccuracy increases as the sphere and the equatorial plane are approached.

In this regard, we will consider below two particular cases when the potential components are calculated in a relatively simple way, which makes it easy to analyze the solution. The first case refers to the region of space near the axis $OZ$, where it can be assumed that $z \approx R$, $R >> x$, $R >> y$. The second case refers to the points on the sphere's surface, where $z = 0$, $R^2 = x^2 + y^2$, while $R = a$.

### 2.9.1. The case when $z \approx R$

In this case, at $x \approx 0$, $y \approx 0$, the integrals $I_{11}$ and $I_{12}$, defined in (64) and found in (67-68), can be simplified, if we make substitution:

$$\frac{1}{\sqrt{R^2 + z_d^2 - 2zz_d + \rho^2 - 2\rho x\cos\phi - 2\rho y\sin\phi}} \approx \frac{1 + \dfrac{\rho x\cos\phi + \rho y\sin\phi}{R^2 + z_d^2 - 2zz_d + \rho^2}}{\sqrt{R^2 + z_d^2 - 2zz_d + \rho^2}}.$$

This gives the following:

$$I_{11} \approx \int_0^{2\pi} \frac{\left(1 + \dfrac{\rho x\cos\phi + \rho y\sin\phi}{R^2 + z_d^2 - 2zz_d + \rho^2}\right)}{\sqrt{R^2 + z_d^2 - 2zz_d + \rho^2}} \cos\phi \, d\phi = \frac{\pi\rho x}{\left(R^2 + z_d^2 - 2zz_d + \rho^2\right)^{3/2}}.$$

$$I_{12} \approx \int_0^{2\pi} \frac{\left(1 + \dfrac{\rho x\cos\phi + \rho y\sin\phi}{R^2 + z_d^2 - 2zz_d + \rho^2}\right)}{\sqrt{R^2 + z_d^2 - 2zz_d + \rho^2}} \sin\phi \, d\phi = \frac{\pi\rho y}{\left(R^2 + z_d^2 - 2zz_d + \rho^2\right)^{3/2}}.$$

Similarly, for the integrals in (82), we use approximate expressions:

$$\frac{1}{\left(R^2 + z_d^2 - 2zz_d + \rho^2 - 2\rho x\cos\phi - 2\rho y\sin\phi\right)^{3/2}} \approx \frac{1 + \dfrac{3\rho x\cos\phi + 3\rho y\sin\phi}{R^2 + z_d^2 - 2zz_d + \rho^2}}{\left(R^2 + z_d^2 - 2zz_d + \rho^2\right)^{3/2}}.$$

$$\frac{1}{R^2 + z_d^2 - 2zz_d + \rho^2 - 2\rho x\cos\phi - 2\rho y\sin\phi} \approx \frac{1 + \dfrac{2\rho x\cos\phi + 2\rho y\sin\phi}{R^2 + z_d^2 - 2zz_d + \rho^2}}{R^2 + z_d^2 - 2zz_d + \rho^2}.$$

$$\frac{1}{\sqrt{R^2+z_d^2-2zz_d+\rho^2-2\rho x\cos\phi-2\rho y\sin\phi}}\approx$$

$$\approx\frac{1+\dfrac{\rho x\cos\phi+\rho y\sin\phi}{R^2+z_d^2-2zz_d+\rho^2}+\dfrac{3\left(\rho x\cos\phi+\rho y\sin\phi\right)^2}{2\left(R^2+z_d^2-2zz_d+\rho^2\right)^2}}{\sqrt{R^2+z_d^2-2zz_d+\rho^2}}.$$

With the help of it we find:

$$I_{18}\approx\int_0^{2\pi}\frac{\left(1+\dfrac{3\rho x\cos\phi+3\rho y\sin\phi}{R^2+z_d^2-2zz_d+\rho^2}\right)\left(x\sin\phi-y\cos\phi\right)^2\sin\phi}{\left(R^2+z_d^2-2zz_d+\rho^2\right)^{3/2}}d\phi=\frac{3\pi\rho y\left(x^2+y^2\right)}{4\left(R^2+z_d^2-2zz_d+\rho^2\right)^{5/2}}.$$

$$I_{19}\approx\int_0^{2\pi}\frac{\left(1+\dfrac{3\rho x\cos\phi+3\rho y\sin\phi}{R^2+z_d^2-2zz_d+\rho^2}\right)\left(x\sin\phi-y\cos\phi\right)^2\cos\phi}{\left(R^2+z_d^2-2zz_d+\rho^2\right)^{3/2}}d\phi=\frac{3\pi\rho x\left(x^2+y^2\right)}{4\left(R^2+z_d^2-2zz_d+\rho^2\right)^{5/2}}.$$

$$I_{20}\approx\int_0^{2\pi}\frac{\left(1+\dfrac{2\rho x\cos\phi+2\rho y\sin\phi}{R^2+z_d^2-2zz_d+\rho^2}\right)\left(x\sin\phi-y\cos\phi\right)^2\sin\phi}{R^2+z_d^2-2zz_d+\rho^2}d\phi=\frac{\pi\rho y\left(x^2+y^2\right)}{2\left(R^2+z_d^2-2zz_d+\rho^2\right)^2}.$$

$$I_{21}\approx\int_0^{2\pi}\frac{\left(1+\dfrac{2\rho x\cos\phi+2\rho y\sin\phi}{R^2+z_d^2-2zz_d+\rho^2}\right)\left(x\sin\phi-y\cos\phi\right)^2\cos\phi}{R^2+z_d^2-2zz_d+\rho^2}d\phi=\frac{\pi\rho x\left(x^2+y^2\right)}{2\left(R^2+z_d^2-2zz_d+\rho^2\right)^2}.$$

$$I_{24}\approx\int_0^{2\pi}\frac{\left(1+\dfrac{\rho x\cos\phi+\rho y\sin\phi}{R^2+z_d^2-2zz_d+\rho^2}+\dfrac{3\left(\rho x\cos\phi+\rho y\sin\phi\right)^2}{2\left(R^2+z_d^2-2zz_d+\rho^2\right)^2}\right)\sin\phi\cos\phi}{\sqrt{R^2+z_d^2-2zz_d+\rho^2}}d\phi=$$

$$=\frac{3\pi\rho^2xy}{4\left(R^2+z_d^2-2zz_d+\rho^2\right)^{5/2}}.$$

As for the integrals $I_{22}$ and $I_{23}$ in (82), we obtain for them $I_{22} \approx I_{23} \approx \frac{\pi}{\sqrt{R^2 + z_d^2 - 2zz_d + \rho^2}}$.

With this in mind, instead of (83) we have:

$$A_{dx}(z \approx R) \approx -\frac{\mu_0 \omega y s \rho_{0q}}{4} \int_0^{\rho_d} \left( \begin{array}{l} \frac{\rho}{\left(R^2 + z_d^2 - 2zz_d + \rho^2\right)^{3/2}} - \\ -\frac{3\omega^2 \rho^3 \left(x^2 + y^2\right)}{8c^2 \left(R^2 + z_d^2 - 2zz_d + \rho^2\right)^{5/2}} + \\ +\frac{3\omega^2 \rho^3 x^2}{4c^2 \left(R^2 + z_d^2 - 2zz_d + \rho^2\right)^{5/2}} + \\ +\frac{\omega^2 \rho}{c^2 \sqrt{R^2 + z_d^2 - 2zz_d + \rho^2}} \end{array} \right) \gamma' \rho^2 d\rho +$$

$$+\frac{\mu_0 \omega^4 x s \rho_{0q}}{4c^3} \int_0^{\rho_d} \left( 1 - \frac{\rho^2 \left(x^2 + y^2\right)}{4\left(R^2 + z_d^2 - 2zz_d + \rho^2\right)^2} \right) \gamma' \rho^3 d\rho.$$

$$A_{dy}(z \approx R) \approx \frac{\mu_0 \omega x s \rho_{0q}}{4} \int_0^{\rho_d} \left( \begin{array}{l} \frac{\rho}{\left(R^2 + z_d^2 - 2zz_d + \rho^2\right)^{3/2}} - \\ -\frac{3\omega^2 \rho^3 \left(x^2 + y^2\right)}{8c^2 \left(R^2 + z_d^2 - 2zz_d + \rho^2\right)^{5/2}} + \\ +\frac{\omega^2 \rho}{c^2 \sqrt{R^2 + z_d^2 - 2zz_d + \rho^2}} + \\ +\frac{3\omega^2 \rho^3 y^2}{4c^2 \left(R^2 + z_d^2 - 2zz_d + \rho^2\right)^{5/2}} \end{array} \right) \gamma' \rho^2 d\rho +$$

$$+\frac{\mu_0 \omega^4 y s \rho_{0q}}{4c^3} \int_0^{\rho_d} \left( 1 - \frac{\rho^2 \left(x^2 + y^2\right)}{4\left(R^2 + z_d^2 - 2zz_d + \rho^2\right)^2} \right) \gamma' \rho^3 d\rho.$$

In these expressions, we can neglect the small terms containing $x^2$, $y^2$ and $x^2 + y^2$ in the numerator. After substituting the Lorentz factor $\gamma'$ from (27), taking into account the expression $\rho_d^2 = a^2 - z_d^2$, we find the following:

$$A_{dx}(z\approx R)\approx -\frac{\mu_0 c^2 ys\rho_{0q}\gamma_c}{\omega\left(x^2+y^2\right)}\left(1-\frac{2\pi\eta\,\rho_0\,z_d^2}{3c^2}\right)I_3-\frac{\mu_0\omega ys\rho_{0q}}{4}D_4-$$
$$-\frac{\mu_0\omega^3 ys\rho_{0q}\gamma_c}{4c^2}I_{31}+\frac{\mu_0\omega^4 xs\rho_{0q}\gamma_c\left(a^2-z_d^2\right)^2}{16c^3}.$$

$$A_{dy}(z\approx R)\approx \frac{\mu_0 c^2 xs\rho_{0q}\gamma_c}{\omega\left(x^2+y^2\right)}\left(1-\frac{2\pi\eta\,\rho_0\,z_d^2}{3c^2}\right)I_3+\frac{\mu_0\omega xs\rho_{0q}}{4}D_4+$$
$$+\frac{\mu_0\omega^3 xs\rho_{0q}\gamma_c}{4c^2}I_{31}+\frac{\mu_0\omega^4 ys\rho_{0q}\gamma_c\left(a^2-z_d^2\right)^2}{16c^3}.$$

The integral

$$I_3=\frac{\omega^2\left(x^2+y^2\right)}{4c^2}\int_0^{\rho_d}\frac{\rho^3\,d\rho}{\left(R^2+z_d^2-2zz_d+\rho^2\right)^{3/2}}$$

was calculated earlier in (28) and then was used in (29), and the quantity $D_4$ was determined in (42). In addition, the following integral is added:

$$I_{31}=\int_0^{\rho_d}\frac{\rho^3\,d\rho}{\sqrt{R^2+z_d^2-2zz_d+\rho^2}}\ .$$

Instead of (85), the vector potential components near the axis $OZ$ become equal to:

$$A_x(z\approx R)\approx -\frac{\mu_0 c^2\rho_{0q}y\gamma_c}{\omega\left(x^2+y^2\right)}\int_{-a}^{a}\left(1-\frac{2\pi\eta\,\rho_0\,z_d^2}{3c^2}\right)I_3\,dz_d-\frac{\mu_0\omega\rho_{0q}y}{4}\int_{-a}^{a}D_4\,dz_d-$$
$$-\frac{\mu_0\omega^3\rho_{0q}y\gamma_c}{4c^2}\int_{-a}^{a}I_{31}\,dz_d+\frac{\mu_0\omega^4\rho_{0q}x\gamma_c}{16c^3}\int_{-a}^{a}\left(a^2-z_d^2\right)^2dz_d.$$

$$A_y(z\approx R)\approx \frac{\mu_0 c^2\rho_{0q}x\gamma_c}{\omega\left(x^2+y^2\right)}\int_{-a}^{a}\left(1-\frac{2\pi\eta\,\rho_0\,z_d^2}{3c^2}\right)I_3\,dz_d+\frac{\mu_0\omega\rho_{0q}x}{4}\int_{-a}^{a}D_4\,dz_d+$$
$$+\frac{\mu_0\omega^3\rho_{0q}x\gamma_c}{4c^2}\int_{-a}^{a}I_{31}\,dz_d+\frac{\mu_0\omega^4\rho_{0q}y\gamma_c}{16c^3}\int_{-a}^{a}\left(a^2-z_d^2\right)^2dz_d.$$

At large distances, when $z \approx R$, $R > a$, we will obtain

$$\int_{-a}^{a} D_4 \, d z_d \approx -\frac{2\pi \eta \rho_0 \gamma_c}{3c^2} \left( \frac{16a^7}{105R^3} - \frac{16a^9}{315R^5} \right), \qquad \int_{-a}^{a} I_{31} \, d z_d \approx \frac{4a^5}{15R} - \frac{4a^7}{105R^3},$$

$$\int_{-a}^{a} \left( 1 - \frac{2\pi \eta \rho_0 z_d^2}{3c^2} \right) I_3 \, d z_d \approx \frac{\omega^2 a^5 (x^2 + y^2)}{15c^2 R^3} - \frac{2\pi \eta \rho_0 \omega^2 (x^2 + y^2)}{12c^4} \left( \frac{4a^7}{105R^3} + \frac{16a^9}{315R^5} \right),$$

and then the following is obtained:

$$A_x (z \approx R > a) \approx -\frac{\mu_0 \omega \rho_{0q} a^5 y \gamma_c}{15R^3} \left( 1 - \frac{10\pi \eta \rho_0 a^2}{21c^2} - \frac{\omega^2 a^2}{7c^2} \right) -$$
$$- \frac{\mu_0 \omega^3 \rho_{0q} a^5 y \gamma_c}{15c^2 R} + \frac{\mu_0 \omega^4 \rho_{0q} a^5 x \gamma_c}{15c^3}.$$

$$A_y (z \approx R > a) \approx \frac{\mu_0 \omega \rho_{0q} a^5 x \gamma_c}{15R^3} \left( 1 - \frac{10\pi \eta \rho_0 a^2}{21c^2} - \frac{\omega^2 a^2}{7c^2} \right) +$$
$$+ \frac{\mu_0 \omega^3 \rho_{0q} a^5 x \gamma_c}{15c^2 R} + \frac{\mu_0 \omega^4 \rho_{0q} a^5 y \gamma_c}{15c^3}.$$

(87)

### 2.9.2. The case when $R = \sqrt{x^2 + y^2} = a$

Let us now consider the second case, referring to the points on the sphere's surface, where $z = 0$, $R^2 = x^2 + y^2$, while $R = a$.

In order to simplify the calculations, in (85) we will limit ourselves to only the largest terms that do not contain $c^2$ and $c^3$ in the denominator. This gives us the following:

$$A_x \left( R = \sqrt{x^2 + y^2} = a \right) \approx -\frac{\mu_0 \omega \rho_{0q} y \gamma_c}{4} \int_{-a}^{a} \left( I_{26} - \frac{9a}{2} I_{29} \right) d z_d \, .$$

$$A_y \left( R = \sqrt{x^2 + y^2} = a \right) \approx \frac{\mu_0 \omega \rho_{0q} x \gamma_c}{4} \int_{-a}^{a} \left( I_{26} - \frac{9a}{2} I_{29} \right) d z_d \, .$$

(88)

The integrals $I_{26}$ and $I_{29}$ in (84), taking into account the relation $\rho_d^2 = a^2 - z_d^2$, will be equal to:

$$I_{26}\left(R=\sqrt{x^2+y^2}=a\right)=\frac{z_d^2+5a\sqrt{a^2-z_d^2}+2a^2}{\sqrt{2a^2+2a\sqrt{a^2-z_d^2}}}-\frac{a^3\left(\sqrt{a^2-z_d^2}+a\right)}{z_d^2\sqrt{2a^2+2a\sqrt{a^2-z_d^2}}}-$$
$$-\frac{2z_d^2+a^2}{\sqrt{a^2+z_d^2}}+\frac{a^4}{z_d^2\sqrt{a^2+z_d^2}}-3a\,\mathrm{Arsh}\frac{\sqrt{a^2-z_d^2}+a}{z_d}+3a\,\mathrm{Arsh}\frac{a}{z_d}.$$

$$I_{29}\left(R=\sqrt{x^2+y^2}=a\right)=-\frac{\left(a+\sqrt{a^2-z_d^2}\right)^3}{3\left(2a^2+2a\sqrt{a^2-z_d^2}\right)^{3/2}}+\frac{4a\left(a+\sqrt{a^2-z_d^2}\right)^2}{3\left(2a^2+2a\sqrt{a^2-z_d^2}\right)^{3/2}}-$$
$$-\frac{2a^2\left(a+\sqrt{a^2-z_d^2}\right)}{\left(2a^2+2a\sqrt{a^2-z_d^2}\right)^{3/2}}+\frac{4a^3}{3\left(2a^2+2a\sqrt{a^2-z_d^2}\right)^{3/2}}-\frac{a^4\left(a+\sqrt{a^2-z_d^2}\right)^3}{3z_d^4\left(2a^2+2a\sqrt{a^2-z_d^2}\right)^{3/2}}+$$
$$+\frac{\left(a+\sqrt{a^2-z_d^2}\right)^3}{3z_d^2\sqrt{2a^2+2a\sqrt{a^2-z_d^2}}}-\frac{4a\left(a+\sqrt{a^2-z_d^2}\right)^2}{3z_d^2\sqrt{2a^2+2a\sqrt{a^2-z_d^2}}}-\frac{2\left(a+\sqrt{a^2-z_d^2}\right)}{3\sqrt{2a^2+2a\sqrt{a^2-z_d^2}}}+$$
$$+\frac{4a}{3\sqrt{2a^2+2a\sqrt{a^2-z_d^2}}}+\frac{2a^2\left(a+\sqrt{a^2-z_d^2}\right)}{z_d^2\sqrt{2a^2+2a\sqrt{a^2-z_d^2}}}+\frac{a^4\left(a+\sqrt{a^2-z_d^2}\right)}{z_d^4\sqrt{2a^2+2a\sqrt{a^2-z_d^2}}}-$$
$$-\frac{\sqrt{a^2-z_d^2}\sqrt{2a^2+2a\sqrt{a^2-z_d^2}}}{3z_d^2}+\frac{a\sqrt{2a^2+2a\sqrt{a^2-z_d^2}}}{z_d^2}-$$
$$-\frac{2a^3}{3z_d^4}\sqrt{a^2+z_d^2}-\frac{5a}{3z_d^2}\sqrt{a^2+z_d^2}+\mathrm{Arsh}\frac{a+\sqrt{a^2-z_d^2}}{z_d}-\mathrm{Arsh}\frac{a}{z_d}.$$

After integration over the variable $z_d$ and substitution into (88), we find:

$$\int_{-a}^{a} I_{26}\,dz_d=-a^2\left[\frac{121\sqrt{2}}{15}+7\ln\mathrm{tg}\left(\frac{\pi}{8}\right)+6\ln\left(\sqrt{2}-1\right)\right]\approx 0{,}04986a^2.$$

$$\int_{-a}^{a}\left(-\frac{9a}{2}I_{29}\right)dz_d=-a^2\left[\frac{2181\sqrt{2}}{80}+\frac{555}{16}\ln\operatorname{tg}\left(\frac{\pi}{8}\right)+9\ln\left(\sqrt{2}-1\right)\right]\approx -0{,}04998a^2 .$$

$$A_x\left(R=\sqrt{x^2+y^2}=a\right)\approx-\frac{10^{-4}\mu_0\omega\rho_{0q}a^2y\gamma_c}{4},$$

$$A_y\left(R=\sqrt{x^2+y^2}=a\right)\approx\frac{10^{-4}\mu_0\omega\rho_{0q}a^2x\gamma_c}{4}. \qquad (89)$$

If we proceed from the form of (60) and (87), the vector potential components at $z=0$ and $R=a$ should be approximately as follows:

$$A_x\approx-\frac{\mu_0\omega\rho_{0q}a^2y\gamma_c}{15}, \qquad A_y\approx\frac{\mu_0\omega\rho_{0q}a^2x\gamma_c}{15}. \qquad (90)$$

Apparently, the difference between the results of (89) and (90) was due to an inaccuracy that arose when the integrals $I_{11}$, $I_{12}$, $I_{18}$, $I_{19}$, $I_{22}$, $I_{23}$, $I_{24}$ were found by expanding the elliptic integrals into a series up to the second-order terms. Although the general behavior of the vector potential outside the rotating charged sphere is determined correctly, but this accuracy turns out to be insufficient for the correct determination of the vector potential directly at the equator of the sphere, and expansion of the elliptic integrals up to higher-order terms is required here.

### 2.10. Electric and magnetic fields in the near zone

According to (46), the electric field depends on the potentials' rates of change in space and time. Since the vector potential at constant rotation of the sphere's particles does not depend on time, then the expression $\mathbf{E}=-\nabla\varphi$ will hold true. Using (72), for the electric field we find the following expression with the use of the sum of integrals:

$$\mathbf{E}\approx-\frac{\rho_{0q}\gamma_c}{4\pi\varepsilon_0}\int_{-a}^{a}\nabla H_1\,dz_d-\frac{\omega^2\rho_{0q}\gamma_c}{8\pi c^2\varepsilon_0}\int_{-a}^{a}\nabla H_3\,dz_d+\frac{\eta\rho_0\rho_{0q}\gamma_c}{6c^2\varepsilon_0}\int_{-a}^{a}z_d^2\nabla H_1\,dz_d+$$
$$+\frac{\eta\rho_0\rho_{0q}\gamma_c}{6c^2\varepsilon_0}\int_{-a}^{a}\nabla H_2\,dz_d .$$

(91)

Still this expression is not final, since in it we must first take the spatial gradients of the quantities $H_1$, $H_2$ and $H_3$ from (71), and then perform integration over the variable $z_d$.

The situation on the axis $OZ$ turns out to be much simpler. Here, in view of (74), the field depends on the distance $R$ approximately according to the Coulomb law for the charge $q_\omega$:

$$\mathbf{E}(z=R) = -\nabla\varphi(z=R) = -\frac{d\varphi(z=R)}{d\mathbf{R}} \approx -\frac{q_\omega}{4\pi\varepsilon_0}\frac{d}{d\mathbf{R}}\frac{1}{\sqrt{\mathbf{R}^2}} = \frac{q_\omega \mathbf{R}}{4\pi\varepsilon_0 R^3}. \quad (92)$$

At small $z$, when $x^2+y^2 \approx R^2$ and $R >> a$, in order to estimate the electric field we can use (79):

$$\mathbf{E}(R>>a) = -\nabla\varphi(R>>a) \approx \frac{q_\omega \mathbf{R}}{4\pi\varepsilon_0 R^3}\left(1+\frac{\omega^2 a^2}{10c^2}\right). \quad (93)$$

If we proceed from the form of (80), then at $z=0$, $R \approx a$, for the electric field we obtain the following:

$$\mathbf{E}(R\approx a) = -\nabla\varphi(R\approx a) \approx \frac{q_\omega \mathbf{R}}{4\pi\varepsilon_0 R^3}\left(1+\frac{6\omega^2 a^2}{c^2}\right). \quad (94)$$

The inaccuracy in the definition of $\mathbf{E}(R\approx a)$ depends on the inaccuracy of the potential in (80).

In (85) approximate expressions were presented for the vector potential components $\mathbf{A}$, while integrals (86) must be substituted into these expressions. The subsequent application of the curl operation allows us to find the magnetic field by the formula $\mathbf{B} = \nabla\times\mathbf{A}$, however the result is cumbersome.

The expressions for the vector potential components are greatly simplified near the axis $OZ$. Leaving the largest terms in (87) and taking into account that $A_z = 0$, we find:

$$A_x(z\approx R>a) \approx -\frac{\mu_0\,\omega\,\rho_{0q}\,a^5 y\gamma_c}{15R^3}\left(1-\frac{10\pi\eta\,\rho_0 a^2}{21c^2}-\frac{\omega^2 a^2}{7c^2}\right).$$

$$A_y(z \approx R > a) \approx \frac{\mu_0\, \omega\, \rho_{0q}\, a^5 x \gamma_c}{15R^3}\left(1 - \frac{10\pi\eta\,\rho_0 a^2}{21c^2} - \frac{\omega^2 a^2}{7c^2}\right).$$

$$B_x(z \approx R > a) = \frac{\partial A_z}{\partial y} - \frac{\partial A_y}{\partial z} \approx \frac{\mu_0\, \omega\, \rho_{0q}\, a^5 x z \gamma_c}{5R^5}\left(1 - \frac{10\pi\eta\,\rho_0 a^2}{21c^2} - \frac{\omega^2 a^2}{7c^2}\right).$$

$$B_y(z \approx R > a) = \frac{\partial A_x}{\partial z} - \frac{\partial A_z}{\partial x} \approx \frac{\mu_0\, \omega\, \rho_{0q}\, a^5 y z \gamma_c}{5R^5}\left(1 - \frac{10\pi\eta\,\rho_0 a^2}{21c^2} - \frac{\omega^2 a^2}{7c^2}\right)$$

$$B_z(z \approx R > a) = \frac{\partial A_y}{\partial x} - \frac{\partial A_x}{\partial y} \approx \frac{\mu_0\, \omega\, \rho_{0q}\, a^5 \gamma_c \left(2R^2 - 3x^2 - 3y^2\right)}{15R^5}\left(1 - \frac{10\pi\eta\,\rho_0 a^2}{21c^2} - \frac{\omega^2 a^2}{7c^2}\right). \tag{95}$$

In (95) $R > a$, but $R$ is not much larger than the sphere's radius $a$.

The components of the magnetic field in (95) actually repeat expressions (47) for the magnetic field in the middle zone, with a slight difference in the terms containing the square of the speed of light.

## 3. Conclusion

The presence of the sphere's rotation leads to addition of cylindrical symmetry about the rotation axis $OZ$ to the sphere's radial symmetry in the formulas for the potential. As a rule, this is expressed in the fact that the scalar potential of the electromagnetic field becomes dependent not only on the sphere's radius $a$, the distance $R$ and the angular velocity $\omega$, but also on the angle $\theta$ between the axis $OZ$ and the direction to the point $P$ where the potential is measured. The latter is confirmed by expressions for the potential (32) in the middle zone, (54) in the far zone, (79) and (80) in the near zone, from which it follows that the potential increases as the radius-vector $\mathbf{R}$ of the observation point approaches the equatorial plane of the rotating sphere. By the order of magnitude, the relative change in the potential does not exceed $\frac{6\omega^2 a^2}{c^2}$, depending on the sphere's radius $a$ and on the angular velocity of rotation $\omega$

.

Thus, for the potential of the rotating sphere we can expect dependence of the form $\varphi = \frac{q_\omega}{4\pi\varepsilon_0 R} F(a, R, \omega, \theta)$, where $F(a, R, \omega, \theta)$ is a certain function. In this case, the remote point $P$, where the potential is calculated, has a radius-vector $\mathbf{R} = (x, y, z) = (R\sin\theta\cos\phi, R\sin\theta\cos\phi, R\cos\theta)$. However, due to the sphere's symmetry, there is no dependence on the angle $\phi$ in the function $F(a, R, \omega, \theta)$ and in the potential $\varphi$.

In addition to the scalar potential, we calculate the vector potential in the middle zone (45), in the far zone (60), and also in the near zone in (85) in view of (86). The first terms in the vector potential components in (45) contain $c^3$ in the denominator, and in (60) the similar terms contain $c$ in the denominator. Such a change in the potential dependence, which appears when going over from the middle zone to the far zone, is a typical consequence of the method of retarded potentials.

In (45) and in (60), there is the same term $-\frac{10\pi\eta\rho_0 a^2}{21c^2}$ associated with the properties of the relativistic uniform system. However, the terms, which are proportional to $\frac{\omega^2 a^2}{c^2}$ and define the dependence on the angular velocity $\omega$, have different coefficients. A similar situation occurs in the near zone for the case when $z \approx R > a$, which is seen in (87).

This can be explained by the fact that in the course of calculations we used not coincident procedures for expansion of functions and their subsequent integration, which give different accuracy. Another possible explanation may be that, indeed, in different zones the dependence on $\omega$ is different. The accuracy of the results obtained can be improved by increasing the terms in expansion of functions into series, however introduction of each new term significantly complicates the calculations. It should be noted that for the purpose of more convenient analytical presentation of the results in an explicit form, some elliptic integrals were expanded into series up to the second- and third-order terms, while other integrals were expanded into series up to the sixth-order terms.

Using the obtained expressions for the scalar and vector potentials, we calculate the electric and magnetic fields outside the rotating charged sphere. The corresponding expressions for the fields are presented in (47) for the middle zone, in (61) for the far zone and in (91) in the near zone for $\mathbf{E}$. The formulas for the electric field $\mathbf{E}$ in the near zone are made more precise in (92) on the axis $OZ$, at small $z$ in (93) and at $z = 0$, $R \approx a$ in (94). In all cases, we can see that the field $\mathbf{E}$ increases due to rotation, while the maximum relative increase reaches the value $\frac{6\omega^2 a^2}{c^2}$ near the sphere's surface in the plane $XOY$.

The components of the magnetic field $\mathbf{B}$ in the near zone on the axis $OZ$ on condition that $z \approx R > a$ are presented in (95). Comparison of (47), (61) and (95) shows that within the framework of the approach used, the obtained approximate expressions for $\mathbf{B}$ differ in different zones in small terms, associated with the dependence on the angular velocity $\omega$, repeating the corresponding difference for the vector potential $\mathbf{A}$.

Due to the charge conservation condition, the charge $q_\omega$ (31) of the rotating sphere is equal to the charge $q_b$ of the fixed sphere in (8). This allows us to equate the Lorentz factor $\gamma_c$ of the particles' motion in the center of the rotating sphere and the similar Lorentz factor $\gamma'_c$ for the same and generally fixed sphere.

The results obtained can be applied to nucleons in atomic nuclei when calculating the binding energy in the gravitational model of strong interaction, which takes into account attraction of nucleons to each other in the strong gravitational field, repulsion of protons due to the electric force, repulsion of nucleons' magnetic moments oriented in the combined magnetic field, as well as interaction of the nucleons' spin gravitational moments in the torsion field of strong gravitation due to the nucleons' proper rotation. Since near the equatorial plane at the surface of a rotating proton the electric potential can be increased due to the addition of the order of $\frac{6\omega^2 a^2}{c^2}$ according to (80), then at a typical angular rotation velocity $\omega = 1.03 \times 10^{23}$ rad/s, according to [16], and at the proton radius of the order of $8.73 \times 10^{-16}$ m, this increases the potential by a factor of 1.54. As a result, this also has an impact on the value of the binding energy of atomic nuclei.

Similar calculation for the neutron star PSR J1614–2230, for which the angular velocity of rotation is $\omega = 1.994 \times 10^3$ rad/s and the radius is $a = 12.8$ km according to [17], gives $\frac{\omega a}{c} = 8.51 \times 10^{-2}$ and $\frac{6\omega^2 a^2}{c^2} \approx 0.04$. So if this star were charged, the field near the star's equator would probably also be increased by a factor of $1 + \frac{6\omega^2 a^2}{c^2} \approx 1.04$ as compared to the field of a non-rotating star. The same applies to the gravitational field in the covariant theory of gravitation, the equations of which are similar to the equations of the electromagnetic field [12].

Due to the fact that the calculations contain a great number of integrals, the key details of these calculations are presented in special files, which are included in an appendix to this work [18].

## References

1. Olbert S. and Belcher J.W. The Electromagnetic Fields of a Spinning Spherical Shell of Charge. arXiv:1010.1917 (2010).
2. Jackson, J.D. Classical Electrodynamics, 2nd Edition (John Wiley and Sons, 1975) Chapter 3 and 5.
3. Griffiths D. J. (2007) Introduction to Electrodynamics, 3rd Edition; Prentice Hall - Problem 5.29.
4. Redzic D.V. Electromagnetostatic charges and fields in a rotating conducting sphere. Progress In Electromagnetics Research, Vol. 110, pp. 83-401 (2010). http://dx.doi.org/10.2528/PIER10100504.
5. Gron O. and Voyenli K. Charge distributions in rotating conductors. European Journal of Physics, Vol. 3, Number 4, pp. 210-214 (1982). https://doi.org/10.1088/0143-0807/3/4/004.
6. Marsh J.S. Magnetic and electric fields of rotating charge distributions. American Journal of Physics, Vol. 50, Issue 1, pp. 51-53 (1982). https://doi.org/10.1119/1.13006.
7. Marsh J.S. Magnetic and electric fields of rotating charge distributions II. American Journal of Physics, Vol. 52, Issue 8, pp. 758-759 (1984). https://doi.org/10.1119/1.13852.
8. Fedosin S.G. The Integral Energy-Momentum 4-Vector and Analysis of 4/3 Problem Based on the Pressure Field and Acceleration Field. American Journal of Modern Physics, Vol. 3, No. 4, pp. 152-167 (2014). https://doi.org/10.11648/j.ajmp.20140304.12.
9. Fedosin S.G. Relativistic Energy and Mass in the Weak Field Limit. Jordan Journal of Physics. Vol. 8, No. 1, pp. 1-16 (2015). http://dx.doi.org/10.5281/zenodo.889210.
10. Fedosin S.G. About the cosmological constant, acceleration field, pressure field and energy. Jordan Journal of Physics. Vol. 9, No. 1, pp. 1-30 (2016). http://dx.doi.org/10.5281/zenodo.889304.
11. Fedosin S.G. The electromagnetic field in the relativistic uniform model. International Journal of Pure and Applied Sciences, Vol. 4, Issue. 2, pp. 110-116 (2018). http://dx.doi.org/10.29132/ijpas.430614.
12. Fedosin S.G. The Gravitational Field in the Relativistic Uniform Model within the Framework of the Covariant Theory of Gravitation. International Letters of Chemistry, Physics and Astronomy, Vol. 78, pp. 39-50 (2018). http://dx.doi.org/10.18052/www.scipress.com/ILCPA.78.39.

13. Jefimenko O.D. The effect of radial acceleration on the electric and magnetic fields of circular currents and rotating charges. J. Phys. A: Math. Gen. Vol. 34, No. 31, pp. 6143-6156 (2001). http://dx.doi.org/10.1088/0305-4470/34/31/309.

14. Healy W.P. Comment on 'The effect of radial acceleration on the electric and magnetic fields of circular currents and rotating charges'. J. Phys. A: Math. Gen. Vol. 35, pp. 2527-2531 (2002). https://doi.org/10.1088/0305-4470/35/10/403.

15. Fedosin S.G. Equations of Motion in the Theory of Relativistic Vector Fields. International Letters of Chemistry, Physics and Astronomy, Vol. 83, pp. 12-30 (2019). https://doi.org/10.18052/www.scipress.com/ILCPA.83.12.

16. Fedosin S.G. The radius of the proton in the self-consistent model. Hadronic Journal, Vol. 35, No. 4, pp. 349-363 (2012). http://dx.doi.org/10.5281/zenodo.889451.

17. Demorest P. B., Pennucci T., Ransom S.M., Roberts M.S.E., Hessels J.W.T. A two-solar-mass neutron star measured using Shapiro delay. Nature. Vol. 467 (7319), pp. 1081-1083 (2010). doi:10.1038/nature09466.

18. Fedosin S.G. Calculation of integrals and related data. Appendix to the article «The electromagnetic field outside the steadily rotating relativistic uniform system». [Data set]. Zenodo (2020). https://doi.org/10.5281/zenodo.4308209.